\documentclass[aps,twocolumn,superscriptaddress,bibnotes,nofootinbib]{revtex4-2}

\usepackage[T1]{fontenc}
\usepackage[utf8]{inputenc}
\usepackage{amsmath,amssymb,mathtools,bm,bbm}
\usepackage{graphicx}
\usepackage{microtype}
\usepackage{physics}
\usepackage{appendix}

\usepackage[colorlinks,
citecolor=teal,
linkcolor=teal,
urlcolor=teal]{hyperref}
\usepackage{orcidlink}

\newcommand{\ii}{\mathrm{i}}
\newcommand{\ee}{\mathrm{e}}
\newcommand{\Z}{\mathbb{Z}}
\newcommand{\FNL}{\mathrm{FNL}}
\newcommand{\NL}{\mathrm{NL}}

\begin{document}

\title{Entanglement spectrum and magic in higher-dimensional \\ free fermionic systems}

\author{Beatrice Magni\orcidlink{0009-0009-8577-0525}}
\email{bmagni@uni-koeln.de}
\thanks{These authors contributed equally to this work}
\affiliation{ Institut für Theoretische Physik, Universität zu Köln, Zülpicher Strasse 77, 50937 Köln, Germany}

\author{Daniele Iannotti\orcidlink{0009-0009-0738-5998}}
\email{d.iannotti@ssmeridionale.it}
\thanks{These authors contributed equally to this work}
\affiliation{ Institut für Theoretische Physik, Universität zu Köln, Zülpicher Strasse 77, 50937 Köln, Germany}
\affiliation{Scuola Superiore Meridionale, Largo S. Marcellino 10, 80138 Napoli, Italy}
\affiliation{INFN Sezione di Napoli, via Cintia, 80126 Napoli, Italy}

\author{Riccardo Travaglino\orcidlink{0000-0002-8768-4462}}
\email{rtravagl@sissa.it}
\thanks{These authors contributed equally to this work}
\affiliation{SISSA and INFN Sezione di Trieste, via Bonomea 265, I-34136 Trieste, Italy}

\date{\today}

\begin{abstract}
We study fermionic non-local magic in higher-dimensional free-fermion systems through the lens of the entanglement spectrum. For translationally invariant Gaussian states in strip geometries, dimensional reduction maps the problem to independent lower-dimensional momentum sectors, leading in two dimensions to a multiplicative logarithmic violation of the boundary law analogous to Gioev-Klich-Widom scaling. We further show that the second Rényi fermionic non-local magic is bounded by the capacity of entanglement and introduce an entanglement-temperature deformation that probes the low-energy structure of the entanglement Hamiltonian. Applied to the matter-Majorana sector of the Kitaev model, this response is exponentially suppressed in the Abelian phase and algebraic in the non-Abelian phase, where its ratio with the capacity approaches a universal constant. Out of equilibrium, we extend the quasi-particle description to fermionic non-local magic in higher-dimensional quenches and show that random Gaussian circuits exhibit diffusive spreading in two spatial dimensions.
\end{abstract}

\maketitle

\section{Introduction}
\label{sec:intro}
Entanglement and magic, or nonstabilizerness, are two fundamental notions of quantum resources that capture complementary aspects of many-body quantum states~\cite{Horodecki2009,Chitambar2019resources,Veitch2014resource}, with distinct operational meanings. Low-entanglement states can often be simulated efficiently by tensor-network methods~\cite{vidal2003,daley2004time,Schollw_ck_2011}, whereas low-magic states are accessible through stabilizer-based techniques such as tableau simulation~\cite{Gottesman1998,Bravyi2019simulationofquantum} or Pauli-propagation methods~\cite{dowling2025magic,rudolph2025paulipropagationcomputationalframework}. While entanglement is by now a central and well-characterized concept in many-body physics~\cite{amico2008entanglement,Calabrese_2009_1, Latorre_2009,arealaws,Laflorencie_2016}, the role of magic has come into focus more recently, driven in part by stabilizer entropies and related scalable probes of nonstabilizerness~\cite{Leone2022SRE,leone2024stabilizer,huang2026fastexactapproachstabilizer,xiao2026exponentiallyacceleratedsamplingpauli,sierant2026computingquantummagicstate,Bittel_2026,Iannotti_2026,cusumano2025nonstabilizernessviolationschshinequalities, Hoshino_2026, White_2021,Bittel_2026,iannotti2026nonstabilizernessu1symmetrychaotic,cepollaro2025stabilizerentropysubspaces}.
A central question is how entanglement and magic combine in extended quantum systems~\cite{tirrito2024quantifying,szombathy,Iannotti2025entanglement}. The total magic of a many-body state contains both local and genuinely non-local contributions~\cite{Oliviero_2022,Viscardi_2026,Turkeshi_2025,odavic2025stabilizer,tirrito2025anticoncentration,falcao2025nonstabilizerness, PhysRevB.106.214316, p8dn-glcw, lio2026quantumstatedesignsmagic, aditya2025growthspreadingquantumresources,turkeshi_pauli_2023, piro2,tarabunga2023manybody,oshino2026stabilizer,hoshino2025stabilizerrenyientropyencodes, aditya2025mpembaeffectsquantumcomplexity, magni2025quantumcomplexitychaosmanyqudit, magni2025anticoncentrationcliffordcircuitsbeyond,fang2026stabilizerrenyimicroscopycriticalcorrelations, Magni_2026, jasser2025stabilizer, EspositoStyliaris2026, jha2026nonstabilizernessentanglement21dimensionalsu2}; while the former is relevant for quantum computation, the latter displays the coarse-grained features relevant for many-body physics.
Non-local magic was introduced to isolate the irreducible nonstabilizerness that cannot be gauged away by local basis changes, thereby separating magic associated with correlations across a bipartition from purely local contributions, and it has since been connected with various concepts such as topological phases, quantum geometry, and holography~\cite{Cao_2025,munizzi2026magicnoncliffordgatestopological,cao2024non, sierant2026exactquantificationnonlocalmagic, cao2026statedependentgeometriesmagicenrichedquantum,biswas2026observationgravitylikesignaturesholographic, liu2026entanglementantiflatnessnonlocalnonstabilizerness,Grieninger_2026}. In generic quantum states, however, this definition involves a minimization over local unitaries on the full Hilbert space and is therefore computationally prohibitive beyond small systems~\cite{qian2025quantumnonlocalnonstabilizerness,ahmad2026experimentaldemonstrationnonlocalmagic,torre2026nonlocalmagicentanglementspectrum,franchini2026schmidtgaugenonlocalmagicrepresentation}, though some improvements have emerged recently~\cite{sierant2026exactquantificationnonlocalmagic,viscardi2026nonlocalmagicclosedformsolution,sierant2026spectralgeometrynonlocalstabilizer,karjula2026universalentanglementembezzlementdivergent,liu2026entanglementantiflatnessnonlocalnonstabilizerness}.

Fermionic Gaussian states~\cite{Surace_2022,bravyi2004lagrangianrepresentationfermioniclinear} provide a special setting in which this obstruction can be bypassed. Restricting the optimization to local Gaussian unitaries leads to fermionic non-local magic (FNL), which admits a closed spectral expression in terms of the eigenvalues of the Majorana covariance matrix restricted to the subsystem~\cite{iannotti2026nonlocalmagicresourcesfermionic,collura2026nonlocalnonstabilizernessfreefermion}.
Thus, FNL is easily computable, in contrast with the total magic, whose evaluation generally requires sampling-based approaches such as Majorana Monte Carlo or perfect sampling~\cite{Bera_2025,collura2026nonstabilizernessfermionicgaussianstates}. This makes FNL a natural quantity for studying the scaling of correlated nonstabilizerness in extended free-fermionic systems.

In this work, we study how the entanglement spectrum~\cite{Chandran_2014} controls fermionic non-local magic in higher-dimensional free-fermion systems, both in equilibrium and out of equilibrium.
This puts FNL on a similar footing to entanglement entropies and the capacity of entanglement~\cite{deBoer2019capacity}, while retaining a distinct way of weighting the entanglement spectrum.
More generally, the structure of the entanglement spectrum and the
associated family of Rényi entropies can reveal properties of quantum
phases that are not captured by a single entanglement entropy, including
symmetry-protected topological order and differential local
convertibility~\cite{Cui_2013}.
We use this structure to determine how FNL scales with spatial dimension, how it resolves low-energy features of the entanglement Hamiltonian, and how it spreads under unitary dynamics.

We first consider translationally invariant Gaussian states in strip geometries. Translation symmetry along the entanglement cut allows the restricted covariance matrix to be decomposed into independent momentum sectors, reducing the higher-dimensional problem to a collection of lower-dimensional ones~\cite{PhysRevB.62.4191}. In two dimensions, this gives the same type of boundary-law violation that appears in the Gioev-Klich-Widom scaling of free-fermion entanglement entropy~\cite{Gioev_2006,Swingle_2010, Wolf_2006}, suggesting the corresponding scaling in arbitrary spatial dimension whenever the Fermi surface has codimension one.
Since entanglement and FNL are both spectral quantities, one can also compare non-local magic with the capacity of entanglement, which measures fluctuations of the modular/entanglement Hamiltonian. We show that the second Rényi FNL is bounded from above by the capacity of entanglement, providing a stronger constraint than the corresponding entropy bounds. This relation motivates an entanglement-temperature deformation of FNL, which probes how different regions of the entanglement spectrum contribute to non-local magic.
Applying this construction to the matter-Majorana sector of the Kitaev model~\cite{Kitaev_2006}, we find that FNL directly distinguishes gapped from gapless entanglement spectra~\cite{YaoQi2010}. Its low-entanglement-temperature response is exponentially suppressed in the Abelian phase and algebraic in the non-Abelian phase. In the latter, the ratio between FNL and the capacity of entanglement approaches a universal constant as the entanglement temperature goes to zero, showing that the two quantities probe the same low-energy entanglement modes with different spectral weights.

We then turn to nonequilibrium dynamics. For quenches generated by number-conserving quadratic Hamiltonians, the spectral representation of FNL allows the standard quasi-particle picture of entanglement growth \cite{calabrese2005evolution,fagotti2008evolution,alba1} to be extended to fermionic non-local magic also in higher dimensions. In strip geometries, the dynamics can be understood in terms of the propagation of occupied fermionic modes with their corresponding velocities, leading to a hydrodynamic description of both the growth and saturation of FNL.

Finally, we consider random fermionic Gaussian circuits. These have been extensively studied in one-dimension~\cite{Swann_2025, dias2021diffusiveoperatorspreadingrandom}, as they show non-generic entanglement dynamics, but less has been said about the two-dimensional case. By exploiting the relation with Clifford-Gaussian ensembles~\cite{paviglianitient2026}, we analyze the spreading of entanglement and of individual Majorana modes. The resulting dynamics is diffusive, with the additional spatial directions modifying the overall scaling.
Since FNL is controlled by the same restricted Majorana correlations and is bounded by the Rényi entanglement entropy, this characterizes the diffusive spreading of FNL under random free-fermion dynamics.

These results support the consistency and applicability of FNL in higher-dimensional systems. Moreover, fermionic non-local magic probes features of the entanglement spectrum beyond those captured by the entanglement entropy, providing complementary information about quantum correlations. In this respect, its role is conceptually similar to that of the capacity of entanglement.
Its equilibrium scaling, its response to low-energy entanglement modes, and its dynamical spreading can all be traced back to the geometry and transport of the underlying single-particle entanglement degrees of freedom.

\section{Methods}

\begin{figure*}[t]
    \centering
    \includegraphics[width=0.98\textwidth]{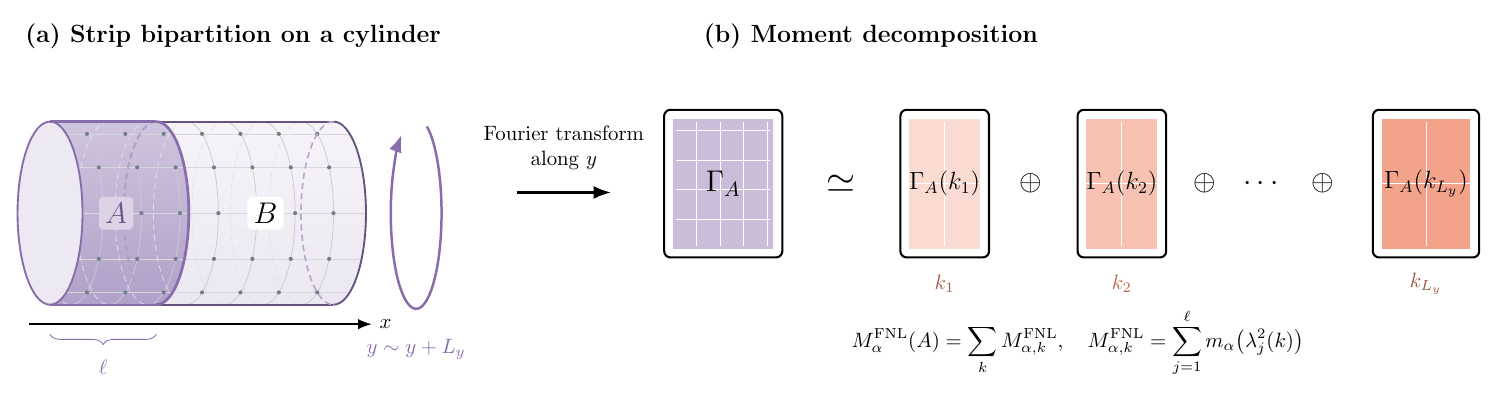}
    \caption{Dimensional reduction of fermionic non-local magic for a pure Gaussian state in two dimensions. (a) The subsystem $A$ (violet) is a strip of width $\ell$ along $x$ that wraps around the cylinder's periodic direction $y$, of circumference $L_y$. (b) For states invariant under translations along $y$, this symmetry is preserved by the restriction to $A$. A transverse Fourier transformation therefore decomposes the restricted Majorana covariance matrix $\Gamma_A$ (violet) into independent momentum blocks $\Gamma_A(k)$ (peach), with $k=2\pi n/L_y$. Each block defines a one-dimensional Gaussian channel. The additive spectral expression for fermionic non-local magic then gives the exact decomposition $M_\alpha^{\FNL}(A)=\sum_k M_{\alpha,k}^{\FNL}$, as expressed in Eq.~\eqref{eq:widom_reduction}.}
    \label{fig:illustration}
\end{figure*}

Consider a lattice of size $L_x\times L_y$, where each lattice position is labeled by the vector $\pmb{r} = (x,y)$, with $x=1,\ldots,L_x$ and $y=1,\ldots,L_y$. Each site hosts a complex fermionic/Dirac mode associated with annihilation and creation operators, $c_{\pmb{r}}$ and $c^\dagger_{\pmb{r}}$, satisfying the canonical anticommutation relations $\{c_{\pmb{r}},c^\dagger_{\pmb{r}'}\}=\delta_{\pmb{r},\pmb{r}'} \mathbb{I}$ and total number of Dirac modes given by $N=L_xL_y$. Equivalently, for every physical site we can introduce real Majorana operators,
\begin{equation}
    \gamma_{2x-1,y} = c_{x,y}+ c^\dagger_{x,y}, \quad \gamma_{2x,y} = i(c_{x,y}-c^\dagger_{x,y}),
    \label{eq:mapping}
\end{equation}
satisfying $\{ \gamma_{\pmb{m}},\gamma_{\pmb{m}'} \}=2\delta_{\pmb{m}\pmb{m'}} \mathbb{I}$ for the Majorana coordinate vector $\pmb{m} = (m_x,m_y)$, with $m_x = 1,..., 2L_x$ and $m_y = 1, ..., L_y$, giving a total number of Majorana modes $2N = (2L_x)L_y$. 
Upon choosing an ordering of the fermionic modes, each Majorana operator can be represented by a Pauli string, via Jordan--Wigner transformation~\cite{Bravyi_2002}. Thus, the Majorana algebra is naturally embedded in the Pauli group $\mathcal{P}_N$ of the corresponding $N$-spin
system, defined as the group generated by the single-spin Pauli operators $\{X_j,Y_j,Z_j\}$ together with the identity, up to overall phases~\cite{Surace_2022}. In the following of the manuscript we will keep the discussion mainly in the Majorana representation.

Fermionic Gaussian unitaries, also known as matchgate unitaries, are characterized by their linear adjoint action on the Majorana operators. In particular, a fermionic Gaussian unitary $U_O\in \mathcal{G}_N$ associated with an orthogonal matrix $O\in\mathrm{O}(2N)$ acts as
\begin{equation}
U_O^\dagger\gamma_{\pmb m}U_O
    =
    \sum_{\pmb m'}O_{\pmb m\pmb m'}\gamma_{\pmb m'}.
\end{equation}

The orthogonality of $O$ follows from the requirement that the Majorana anticommutation relations remain invariant under the transformation.

A pure fermionic Gaussian state, also referred to as a free-fermion state, is obtained by applying a fermionic Gaussian unitary to the fermionic vacuum $\ket{\Psi_O} =U_O \ket{\boldsymbol{0}}$.
Using the present indexing convention, the vacuum density matrix is $|\mathbf{0}\rangle\langle\mathbf{0}| = 2^{-N}\prod_{y=1}^{L_y}\prod_{x=1}^{L_x}(\mathbb{I}-\ii\gamma_{2x-1,y}\gamma_{2x,y})$.
For a generic pure fermionic Gaussian state $\ket{\Psi}$, the two-point correlations can be equivalently encoded in the Dirac and Majorana representations. In the Dirac representation, we define the correlation matrix as
\begin{equation}
\label{eq:corrmat}
    C_{\pmb{r},\pmb{r}'} = \bra{\Psi}c_{\pmb{r}}^\dagger c_{\pmb{r}'}\ket{\Psi}\,,
\end{equation}
while in the Majorana representation the covariance matrix is defined as
\begin{equation}
    \Gamma_{\pmb{m},\pmb{m}'}=i\frac{\langle \Psi |[\gamma_{\pmb{m}},\gamma_{\pmb{m}'}]|\Psi\rangle}{2},
\end{equation}
which is real and antisymmetric. For the fermionic vacuum, the covariance matrix takes the form
\begin{equation}
    \Gamma_0= { \bigoplus_{m=1}^N \begin{pmatrix}
        0 &1 \\-1&0
    \end{pmatrix}}\;,
    \label{eq:gamma_zero}
\end{equation}
and under the action of $U_O$, it transforms according to $\Gamma=O\Gamma_0O^T$.

\subsection{Entanglement entropies and fermionic non-local magic}
Given a pure state $\ket{\Psi}\in \mathcal{H}$, with $\mathcal{H}= \mathcal{H}_A \otimes \mathcal{H}_B$ a decomposition imposed by local obeservables~\cite{Zanardi_2004}, one can define the reduced density matrix on the smaller subsystem as $\rho_A=\Tr_B[\ket{\Psi}\bra{\Psi}]$.
For a pure Gaussian state, the entanglement entropy~\cite{amico2008entanglement} reads as
\begin{equation}
S_\alpha(A)=\frac{1}{1-\alpha}\log \Tr(\rho_A^\alpha)= \sum_j^{|A|} s_\alpha(\lambda_j),
\label{eq:EE}
\end{equation}
with 
\begin{equation}
\label{eq:entropycontribution}
    s_\alpha(x)= \frac{1}{1-\alpha} \log\!\left[
\left(\frac{1-x}{2}\right)^\alpha+\left(\frac{1+x}{2}\right)^\alpha
\right],
\end{equation}
and $\{\pm\lambda_j\}$ are the eigenvalues of $\ii\Gamma_A$, with $\Gamma_A$ the Majorana covariance matrix restricted to $A$.

On the other side, one can evaluate the magic content of the full state through the stabilizer Rényi entropy (SRE)~\cite{Leone2022SRE} as
\begin{equation}
    M_\alpha(\ket{\Psi})= \frac{1}{1-\alpha} \log \left( \sum_{P \in \mathcal{P}_N} \frac{\bra{\Psi}P\ket{\Psi}^{2\alpha}}{d}\right),
    \label{eq:SRE}
\end{equation}
which can also be expressed directly in terms of the covariance matrix~\cite{collura2026nonstabilizernessfermionicgaussianstates}.
Exploiting that Pauli strings are in one-to-one correspondence, up to irrelevant phases, with Majorana monomials $\gamma_I=\prod_{m\in I}\gamma_m$, $I\subseteq[2N]$ (with $[m]:=\{1,\dots,m\}$), we have that
\begin{equation}
       M_\alpha(\ket{\Psi})= \frac{1}{1-\alpha} \log \left( \sum_{I\subseteq[2N]} \frac{\det(\Gamma_I)^{\alpha}}{d}\right),
    \label{eq:SRE_cov}
\end{equation}
 where $\Gamma_I$ is the principal submatrix of $\Gamma$ on $I$.

However, in many-body systems, one is typically interested in universal and emergent properties that are insensitive to microscopic details, in contrast to the fine-grained information captured by Eq.~\eqref{eq:SRE} in the context of quantum computation. One way to address this obstacle when studying both entanglement and magic is to introduce the concept of non-local magic, which describes the irreducible non-stabilizerness tied to bipartite entanglement. In mathematical terms it is defined as 
\begin{equation}
    M_\alpha^\textup{NL}:= \min_{U_A , U_B} M_\alpha\left[(U_A\otimes U_B)|\Psi\rangle\right]\;,\label{eq:defNLmagic}
\end{equation}
with the minimization over factorized unitaries $\mathrm{U}(\mathcal{H}_A)\otimes \mathrm{U}(\mathcal{H}_B)$~\cite{Cao_2025}.
This notion is tied to (but distinct from) that of long-range magic, that is, the non-stabilizerness that cannot be removed by finite-depth local unitary circuits~\cite{Korbany_2025,korbany2026longrangenonstabilizernesstopologicallyencoded, Ellison_2021}.
Even though Eq.~\eqref{eq:defNLmagic} has a conceptually clear meaning, its actual computation is exponentially hard, and it is infeasible beyond a few qubits. 

Hence, for free fermions, one can define the fermionic non-local magic (FNL)~\cite{iannotti2026nonlocalmagicresourcesfermionic,collura2026nonlocalnonstabilizernessfreefermion} which restricts the minimization to local Gaussian unitaries and, in doing so, reduces its expression to a function of the eigenvalues of $i\Gamma_A$:
\begin{equation}
\begin{split}
    M_\alpha^{\FNL}(A)&=\min_{U_{O_A} , U_{O_B}} \!\!M_\alpha \left[ (U_{O_A}\otimes U_{O_B})|\Psi\rangle\right]    \\
    &=\sum_j^{|A|} m_\alpha(\lambda_j^2),
\end{split}
\label{eq:FNLformula}
\end{equation}
with
\begin{equation}
m_\alpha(x)=\frac{1}{1-\alpha}\log\!\left[
\frac{(1-x)^\alpha+1+x^\alpha}{2}
\right],
\quad \alpha\ge2 .
\label{eq:weight}
\end{equation}
This result does not depend on the ordering of the modes in the subsystem A and B~\cite{iannotti2026nonlocalmagicresourcesfermionic}.
In general, the following relationship between the entanglement and FNL~\cite{Cao_2025} holds
\begin{equation}
    M_2^{\FNL}(A)  \leq 
    \begin{cases}
        \,2 S_2(A) \\
        \,S_1(A)
    \end{cases},
    \label{eq:M_S}
\end{equation}
due to the form of Eq.~\eqref{eq:weight}
and Eq.~\eqref{eq:entropycontribution}.
Additionally, a state can be maximally entangled yet carry zero non-local nonstabilizerness across the bipartition, e.g. Bell pairs or rainbow states~\cite{iannotti2026nonlocalmagicresourcesfermionic}.

\subsection{Fermionic non-local magic and capacity of entanglement}
The fact that both entanglement entropies and fermionic non-local magic are spectral functions of the restricted covariance matrix makes it natural to compare FNL with another spectral quantity, which was first connected with non-local magic and gravitational back reaction~\cite{Cao_2025}: the capacity of entanglement.
Originally introduced in the study of the entanglement spectrum of the Kitaev model~\cite{YaoQi2010} it was subsequently developed as a general diagnostic of fluctuations of the modular/entanglement Hamiltonian~\cite{deBoer2019capacity,Arias2023,arias2023b,travaglino2026transport}.

Given a reduced density matrix $\rho_A$, the entanglement Hamiltonian is defined as $K_A=-\log\rho_A$~\cite{EH,ehFF1,EHFF2}. The capacity of entanglement is its variance,
\begin{equation}
C_E(A)=\mathrm{Tr}(\rho_A K_A^2)-\left[\mathrm{Tr}(\rho_A K_A)\right]^2,
\label{eq:CE}
\end{equation}
or equivalently, $C_E(A)=\left.\partial_\alpha^2\log\mathrm{Tr}(\rho_A^\alpha)\right|_{\alpha=1}$.
Thus, while the von Neumann entropy is the expectation value of the entanglement Hamiltonian, the capacity measures its fluctuations, i.e. it provides a measure of spectral antiflatness~\cite{jasser2026journeyflatlanddoesantiflatness}.

For free fermionic systems, the capacity of entanglement is additive over the entanglement Majorana modes, as for the other spectral functions,
\begin{equation}
C_E(A)=\sum_{j}^{|A|}c(\lambda_j),
\label{eq}
\end{equation}
where
\begin{equation}
c(x)=\frac{1-x^2}{4}\left[\log\left(\frac{1-x}{1+x}\right)\right]^2.
\label{eq}
\end{equation}
This expression allows us to compare it with fermionic non-local magic, and find that $m_2(\lambda^2)\leq c(\lambda)$ for $\lambda \in [0,1]$.
It follows that any pure fermionic Gaussian state satisfies 
\begin{equation}
M_2^{\FNL}(A)\leq C_E(A),
\label{eq}
\end{equation}
which consists of a tighter upper bound than Eq.~\eqref{eq:M_S}.
A similar relationship between the capacity of entanglement and a different measure of non-local magic, introduced in~\cite{sierant2026exactquantificationnonlocalmagic}, has also been discussed in a recent work~\cite{aditya2026nonlocalmagicspreadingmanybody} on one-dimensional free fermionic systems.

\section{Gioev-Klich-Widom scaling}

In this section we focus on a cylinder geometry,
\begin{equation}
\label{eq:cylinder}
\Lambda=\{1,\ldots,L_x\}\times\Z_{L_y}
\end{equation}
with periodic boundary conditions along \(y\) and arbitrary boundary conditions along \(x\). The subsystem of interest is the transverse strip $A=\{1,\ldots,\ell\}\times\Z_{L_y}$.
Since the subsystem contains the entire transverse direction, translation invariance along $y$ is preserved by the restriction to $A$.

Let $c_{x,y}$ be the complex fermionic annihilation operator at the lattice site $(x,y)$. One can Fourier transform it along the transverse direction, 
\begin{equation} 
c_{x,k} = \frac{1}{\sqrt{L_y}} \sum_{y=1}^{L_y} \ee^{-\ii ky}c_{x,y}, \qquad k=\frac{2\pi n}{L_y}. 
\end{equation}
The restricted covariance matrix becomes block diagonal in $k$~\cite{Murciano_2020}, 
\begin{equation} 
\Gamma_A = \bigoplus_k \Gamma_A(k). 
\label{eq:widom_block} 
\end{equation} 

Since $M_\alpha^{\FNL}$ depends only on the eigenvalues of the restricted covariance matrix Eq.~\eqref{eq:widom_block}, Eq.~\eqref{eq:FNLformula} becomes 
\begin{equation} 
M_\alpha^{\FNL}(A) = \sum_k M_{\alpha,k}^{\FNL}, \;\; M_{\alpha,k}^{\FNL} = \sum_j m_\alpha\!\left(\lambda^2_j(k)\right), \label{eq:widom_reduction} 
\end{equation} 
where $\{\pm\lambda_j(k)\}$ are the eigenvalues of $\ii\Gamma_A(k)$, see Fig.~\ref{fig:illustration}. Thus the two-dimensional problem is reduced to a collection of one-dimensional Gaussian problems, as for both entanglement entropy and its capacity.

If the one-dimensional problem at fixed $k$ is gapped, its contribution remains finite for $\ell\to\infty$, $M_{\alpha,k}^{\FNL}=O(1)$. 
Summing over the $L_y$ transverse momenta then gives the usual boundary law $M_\alpha^{\FNL}(A)=O(L_y)$.
The situation changes when a finite fraction of the transverse momenta corresponds to critical one-dimensional chains. In this case, each critical channel contributes logarithmically, $M_{\alpha,k}^{\FNL} = \beta_\alpha^{\FNL}(k)\log\ell+O(1)$, where $\beta_\alpha^{\FNL}$ is the one-dimensional critical coefficient. 
Therefore, 
\begin{equation} 
M_\alpha^{\FNL}(A) = \frac{L_y}{2\pi} \log\ell \int_{\text{critical}} dk\, \beta_\alpha^{\FNL} (k)  + O(L_y). \label{eq:widom_general} 
\end{equation}
Hence a finite-measure set of critical transverse momenta produces a multiplicative logarithmic violation of the boundary law, 
\begin{equation} 
M_\alpha^{\FNL}(A) \sim L_y\log\ell. 
\label{eq:widom_LlogL} 
\end{equation} 
For an approximately square geometry, $L_y\sim\ell\sim L$, this becomes $M_\alpha^{\FNL}(A)\sim L\log L$.
The origin of this logarithm is the same dimensional-reduction mechanism underlying Gioev-Klich-Widom scaling of free-fermion entanglement entropy. 
A transverse momentum contributes to the logarithm precisely when the corresponding one-dimensional cut through momentum space intersects the Fermi surface. The measure of the critical values of $k$ is therefore fixed geometrically by the projection of the Fermi surface onto the transverse momentum direction.
Additionally, the full non-local magic in Eq.~\eqref{eq:defNLmagic} is expected to scale as $M_\alpha^{\NL}(A) \sim L_y \log \log \ell$ in analogy with the one-dimensional cases~\cite{sierant2026spectralgeometrynonlocalstabilizer,franchini2026schmidtgaugenonlocalmagicrepresentation}, see Appendix~\ref{ap:Comparison}.

\subsection{Nearest-neighbour hopping Hamiltonian}
\begin{figure}
    \centering
    \includegraphics[width=0.9\linewidth]{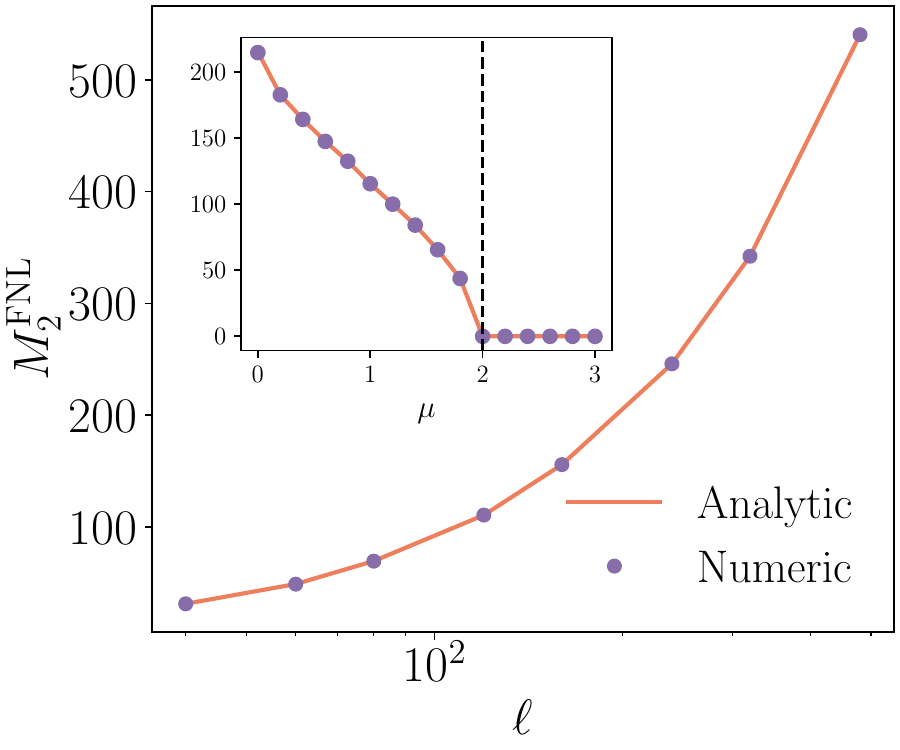}
    \caption{Comparison between Eq.~\eqref{eq:FNL_square_lattice_scaling} and the numerical values obtained directly from the correlation matrix. The strips considered have dimension $L_y = \ell = \{40, 60, 80, 120, 160, 240, 320, 480\}$ with $\mu = 0.5$. In the inset, we show the same comparison of the FNL with fixed $L_y = \ell = 160$ and $\mu \in [0,3]$ varying inside and outside the critical regime (black dashed line).}
    \label{fig:fnl_hopping}
\end{figure}

As a concrete example, consider a two-dimensional square lattice with
isotropic hopping between nearest-neighbour sites
\begin{equation}
H=- \frac{1}{2}\sum_{\langle\bm r,\bm r'\rangle}
(c_{\bm r}^\dagger c_{\bm r'}+ c_{\bm r'}^\dagger c_{\bm r})
+ \mu\sum_{\bm r}c_{\bm r}^\dagger c_{\bm r},
\label{eq:hopping}
\end{equation}
where $\mu$ is the chemical potential for fermions $c_{\bm r}$, with ${\bm r}=(x,y)$ a vector identifying a given lattice site and $\langle\bm r,\bm r'\rangle$ stands for nearest neighbours.
After Fourier transformation along \(y\),
\begin{equation}
H=\sum_k\left[
-\frac{1}{2}\sum_x(c_{x,k}^\dagger c_{x+1,k}+\mathrm{h.c.})
+ \mu_k\sum_x c_{x,k}^\dagger c_{x,k}
\right],
\end{equation}
with $\mu_k=\mu-\cos k$~\cite{Murciano_2020}.
Thus, each transverse momentum defines a one-dimensional tight-binding chain, and after a Jordan-Wigner transformation it is mapped to the spin-1/2 XX chain in a magnetic field. The channel (1D tight-binding model) is gapless exactly when its effective chemical potential lies inside the one-dimensional band $|\mu_k|<1$, which is satisfied for \footnote{For a finite number of chains and $\mu=0$, $n=0$ must also be removed from $\Omega_n$~\cite{Murciano_2020}.}
\begin{equation}
\begin{split}
     n \in \Omega_n &=\Big[0, \frac{\arccos(\mu-1) L_y}{2 \pi} \Big[ \;\cup\\
     &\Big] L_y \left(1-\frac{\arccos(\mu-1) }{2 \pi}\right), L_y-1\Big]\,.
\end{split}
\end{equation}
Similarly to what was done in Ref.~\cite{Jin_2004} for the entanglement entropy, we derive the scaling behavior of $\FNL$ in Appendix~\ref{eq:scalingMFNL}, obtaining
\begin{equation}
\begin{split}
M_2^{\mathrm{FNL}}(A)
&= 2 L_y \,f_{L_y}(\mu) [\beta_2^{\FNL} (\log(2 \ell)+1 +\gamma_E)\\
&- \mathcal{J}_2/ \pi^2]+ \beta_2^{\FNL} A_{L_y}(\mu)\,.
\end{split}
\label{eq:FNL_square_lattice_scaling}
\end{equation}
where $f_{L_y}(\mu)$ is the fraction of critical modes, $\beta_2^{\FNL}$ is the constant coefficient at criticality for the Ising model~\cite{iannotti2026nonlocalmagicresourcesfermionic}, $\gamma_E$ is the Euler constant,  $A_{L_y}(\mu) = \sum_{n\in \Omega_n}\log[1-\left(\mu-\cos\left(\frac{2\pi n}{L_y}\right)\right)^2]$, and 
$\mathcal{J}_2$ is a constant whose expression is given in Appendix~\ref{App:const_Ising}. We show this behavior in Fig.~\ref{fig:fnl_hopping} as a function of the system size and of the chemical potential $\mu$.
The leading growth of the fermionic non-local magic is entirely controlled by the number of gapless transverse modes, yielding an $L_y\log \ell$ scaling. Instead, outside of the critical regime, the FNL vanish completely. 
This is consistent with the analysis in Ref.~\cite{iannotti2026nonlocalmagicresourcesfermionic} of the one-dimensional XY model, which indeed reduces to the Hopping model for certain parameter values. 

\section{Fermionic non-local magic at finite entanglement temperature}

The capacity of entanglement in Eq.\eqref{eq:CE} was originally introduced in the Kitaev
honeycomb model as a probe of the low-energy structure of the
entanglement Hamiltonian \cite{YaoQi2010}. In particular, the capacity
distinguishes the Abelian and non-Abelian phases through the qualitatively
different low-temperature behavior associated with a gapped or gapless
entanglement spectrum. Motivated by this construction, we introduce an
analogous entanglement-temperature deformation of fermionic non-local
magic.

Consider a pure bipartite fermionic Gaussian state with restricted
covariance-matrix eigenvalues $\{\pm\lambda_j\}$. Up to local Gaussian
unitaries, its Gaussian Schmidt decomposition can be written as a product
of entangled fermionic pairs with probabilities
$(1\pm\lambda_j)/2$. One associates with each pair the single-particle
entanglement energy
\begin{equation}
\varepsilon_j
=
\log\left(\frac{1+\lambda_j}{1-\lambda_j}\right)
=
2\operatorname{arctanh}\lambda_j.
\label{eq:entanglement_energy}
\end{equation}
Following the thermal interpretation of the entanglement Hamiltonian, we
introduce an entanglement temperature $T_E$ through
\begin{equation}
\rho_A(T_E)
=\frac{e^{-K_A/T_E}}{\mathrm{Tr}[e^{-K_A/T_E}]},
\label{eq:rho_entanglement_temperature}
\end{equation}
with $K_A=-\log \rho_A$.
The covariance singular values of this state are
\begin{equation}
\lambda_j(T_E)
=
\tanh\left(\frac{\varepsilon_j}{2T_E}\right)
=
\tanh\left[
\frac{\operatorname{arctanh}\lambda_j}{T_E}
\right].
\label{eq:lambda_entanglement_temperature}
\end{equation}
Hence, the pure state associated with these eigenvalues can be written, even though not uniquely, as
\begin{equation}
\small
|\psi(T_E)\rangle
=
\bigotimes_j
\sqrt{\frac{1+\lambda_j(T_E)}{2}}\,
|00\rangle_j
+
\sqrt{\frac{1-\lambda_j(T_E)}{2}}\,
|11\rangle_j.
\end{equation}
The temperature-resolved fermionic non-local magic is then defined as
\begin{equation}
M_\alpha^{\FNL}(T_E)
=
\sum_j
m_\alpha\left(\lambda_j^2(T_E)\right).
\label{eq:FNL_entanglement_temperature}
\end{equation}
At $T_E=1$ the original state is recovered, and therefore $M_\alpha^{\FNL}(T_E=1)=M_\alpha^{\FNL}$.
Notice also that $M_\alpha^{\FNL}(T_E)$ vanishes both for
$T_E\rightarrow0,\infty$. In the first limit, all
finite-entanglement-energy modes become disentangled, while in the second
limit the Schmidt spectrum becomes flat.
In contrast, the capacity evaluated at the same entanglement temperature is
\begin{equation}
C_E(T_E)
=
\sum_j
\frac{(\varepsilon_j/T_E)^2}
{4\cosh^2\left(\varepsilon_j/2T_E\right)}.
\label{eq:capacity_entanglement_temperature}
\end{equation}
Since Eq.~\eqref{eq:lambda_entanglement_temperature} simply produces
another pure fermionic Gaussian state, the pointwise inequality derived
above applies at every $T_E$
\begin{equation}
M_2^{\FNL}(T_E)
\leq
C_E(T_E).
\label{eq:FNL_capacity_temperature_bound}
\end{equation}

\subsection{Matter entanglement Hamiltonian of the Kitaev honeycomb model}

\begin{figure}[t]
    \centering
    \includegraphics[width=
    \linewidth]{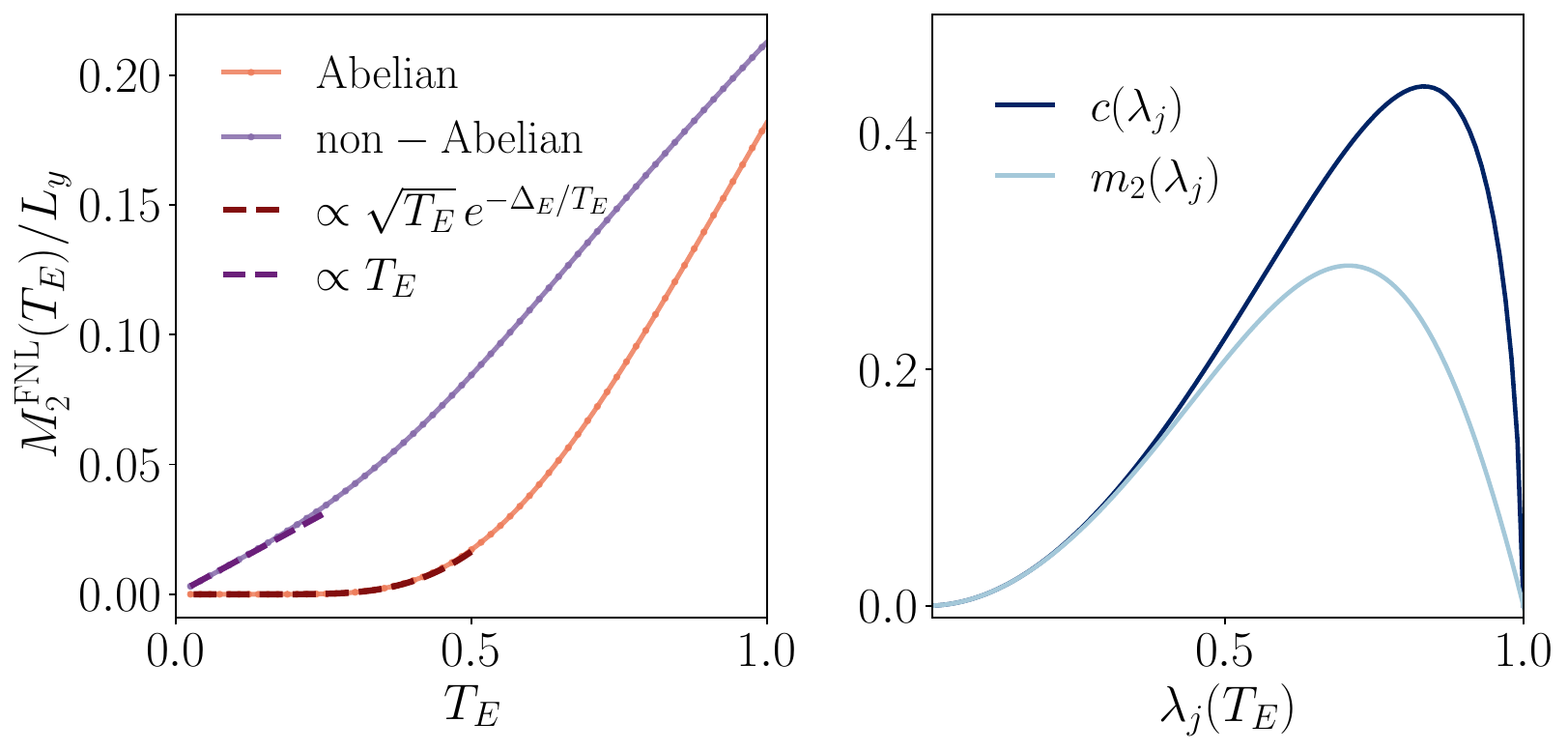}
    \caption{ FNL variation with the entanglement temperature of the matter-Majorana entanglement spectrum in the Kitaev model.
    The left panel shows the temperature-resolved fermionic non-local magic $M_2^{\FNL}(T_E)/L_y$ as a function of the entanglement temperature $T_E$. In the non-Abelian phase the low-temperature behavior is linear, $M_2^{\FNL}(T_E)\propto T_E$, whereas in the Abelian phase it is exponentially suppressed according to $M_2^{\FNL}(T_E)\sim \sqrt{T_E}\exp(-\Delta_E/T_E)$. The dashed lines denote the corresponding low-temperature fits.
    Numerics are performed on a system with $L_x=48$, a subsystem $\ell_x=L_x/2=24$, and $L_y=401$. We use the parameter sets $J_x=J_y=J_z=1$, $J'=0.2$ for the non-Abelian phase and $J_x=J_z=1$, $J_y=2.5$, $J'=0.2$ for the Abelian phase.
    The right panel represents the single-mode behavior of the FNL and the capacity of entanglement. The capacity provides a tighter bound over the FNL, having a similar shape with the same minima but slightly shifted maxima. This was also observed in Ref.~\cite{jasser2026journeyflatlanddoesantiflatness} between the antiflatness measures and the capacity.}
    
    \label{fig:kitaev_temperature_fnl}
\end{figure}

We now study Eq.~\eqref{eq:FNL_entanglement_temperature} in the Kitaev model~\cite{Kitaev_2006} considered in Ref.~\cite{YaoQi2010},
\begin{equation}
\small
H_K
=
-J_x\sum_{\langle i,j\rangle_x}
\sigma_i^x\sigma_j^x
-J_y\sum_{\langle i,j\rangle_y}
\sigma_i^y\sigma_j^y
-J_z\sum_{\langle i,j\rangle_z}
\sigma_i^z\sigma_j^z,
\label{eq:kitaev_honeycomb}
\end{equation}
supplemented by a time-reversal-breaking term expressed as a three-spin interaction of strength $J'$, which introduces a non-Abelian phase while maintaining integrability~\cite{PhysRevLett.99.196805}.
We will study the model in the Majorana representation, where for every fixed
$\mathbb{Z}_2$ gauge configuration, the model reduces to a quadratic
Majorana Hamiltonian.

Importantly, the reduced density matrix of the physical Kitaev state
factorizes at the level of its moments into a gauge contribution and a
free-Majorana contribution~\cite{YaoQi2010}. The gauge contribution has a
flat entanglement spectrum, while the nontrivial low-energy structure of
the entanglement Hamiltonian is entirely contained in the Majorana
sector. In the following, $M_\alpha^{\FNL}$ denotes the fermionic
non-local magic of this Gaussian Majorana sector. Although this quantity
is not, in general, identical to the non-local magic obtained by
optimizing directly in the full physical spin Hilbert space, it still
probes properties of the ground state~\cite{lamma2026magickitaevspinliquids}. 

We consider the torus geometry bipartitioned by two boundaries parallel
to the $y$ direction, as in Ref.~\cite{YaoQi2010}. Translation invariance
along the cut allows the Majorana problem to be decomposed into transverse
momenta $k$. Let $\xi_n(k)$ denote the eigenvalues of the
single-particle correlation matrix used to characterize the
entanglement spectrum. The corresponding positive covariance-matrix
eigenvalues in our notation are
\begin{equation}
\lambda_n(k)
=2\xi_n(k)-1,
\label{eq:kitaev_lambda_xi}
\end{equation}
and the positive single-particle entanglement energies are
\begin{equation}
\varepsilon_n(k)
=
\log\left( \frac{\xi_n(k)}{1-\xi_n(k)}\right)
=
2\operatorname{arctanh}\lambda_n(k).
\label{eq:kitaev_entanglement_energy}
\end{equation}
Consequently, the temperature-resolved FNL can be evaluated directly from
the entanglement spectrum, introducing
\begin{equation}
f_2(x) = -\log\left[ 1-\tanh^2\left(\frac{x}{2}\right)
+\tanh^4\left(\frac{x}{2}\right) \right],
\end{equation}
then 
\begin{equation}
M_2^{\FNL}(T_E) =\sum_{n,k} f_2\left(\varepsilon_n(k)/T_E\right)\,.
\label{eq:kitaev_FNL_temperature}
\end{equation}
We compare the Abelian and non-Abelian topological phases of the model, focusing on their distinct free fermionic entanglement spectra: gapped in the Abelian phase and gapless in the non-Abelian phase. We now show how this distinction is encoded in \(M_2^{\mathrm{FNL}}(T_E)\).

\subsubsection{Abelian phase}

Let $\Delta_E>0$ denote the single-particle entanglement gap, $\varepsilon_n(k) \equiv \varepsilon \geq \Delta_E$.
For $\varepsilon/T_E\gg1$ one has
\begin{equation}
\tanh^2\left(\frac{\varepsilon}{2T_E}\right)
=
1-4\ee^{-\varepsilon/T_E}
+
O\left(\ee^{-2\varepsilon/T_E}\right).
\end{equation}
It follows that
\begin{equation}
m_2\left(\lambda^2(T_E)\right)
=
4\ee^{-\varepsilon/T_E}
+
O\left(\ee^{-2\varepsilon/T_E}\right).
\label{eq:FNL_kernel_large_energy}
\end{equation}
Therefore the low-temperature FNL is exponentially suppressed,
\begin{equation}
M_2^{\FNL}(T_E)
=
O\left(
\ee^{-\Delta_E/T_E}
\right),
\qquad
T_E\rightarrow0.
\label{eq:FNL_abelian_exponential}
\end{equation}
If the lowest entanglement band has a generic quadratic minimum,
$\varepsilon(k)=\Delta_E+a(k-k_0)^2+\cdots$, the momentum integral
gives the more precise scaling (see Appendix \ref{app:quadratic-minimum})
\begin{equation}
M_2^{\FNL}(T_E)
\sim
L_y\sqrt{T_E}\,
\ee^{-\Delta_E/T_E}.
\label{eq:FNL_abelian_asymptotic}
\end{equation}
Thus the temperature-resolved FNL becomes exponentially small in the
Abelian phase, directly reflecting the finite gap of the entanglement
Hamiltonian.

\subsubsection{Non-Abelian phase}

In the non-Abelian phase, the entanglement spectrum contains protected
gapless branches crossing $\xi=1/2$, or equivalently
$\varepsilon=0$. Close to a crossing $k=k_a$, the entanglement energy
is linear,
\begin{equation}
\varepsilon_a(k)
=
v_a|k-k_a|
+
O\left((k-k_a)^2\right).
\label{eq:kitaev_linear_entanglement_branch}
\end{equation}
The low-temperature behavior is therefore controlled entirely by an
$O(T_E)$ window around the gapless points. 
In the continuum limit, along the boundary, it gives
\begin{equation}
M_2^{\FNL}(T_E) \simeq \frac{L_yT_E}{\pi} \sum_a \frac{\mathcal I_2}{v_a},
\label{eq:FNL_nonabelian_linear}
\end{equation}
where $\mathcal I_2 = \int_0^\infty du\,f_2(u) =\frac{\pi^2}{4} -\log^2\left(2+\sqrt{3}\right)$.
Consequently,
\begin{equation}
M_2^{\FNL}(T_E) \propto L_yT_E, \qquad T_E\rightarrow0,
\label{eq:FNL_nonabelian_scaling}
\end{equation}
in direct contrast with the behavior of
Eq.~\eqref{eq:FNL_abelian_exponential}.

The same low-energy modes determine the capacity of entanglement, where for a
linear branch, its kernel gives $\frac{\pi^2}{6}$.
The dependence on the boundary length, entanglement velocity, and the number
of gapless branches therefore cancels in the ratio between the two quantities. Thus, we obtain the universal low-temperature relation
\begin{equation}
\kappa= \lim_{T_E\rightarrow0}
\frac{M_2^{\FNL}(T_E)}
{C_E(T_E)}
=
\frac{6}{\pi^2}
\left[
\frac{\pi^2}{4}
-\log^2\left(2+\sqrt{3}\right)
\right]\,,
\label{eq:FNL_capacity_universal_ratio}
\end{equation}
which is independent of the velocity of the gapless entanglement mode and of the number of equivalent linear branches.

The temperature-resolved FNL therefore reproduces the qualitative
distinction originally detected through the capacity of entanglement,
\begin{equation}
M_2^{\FNL}(T_E)
\sim
\begin{cases}
L_y\sqrt{T_E}\,\ee^{-\Delta_E/T_E},
& \text{Abelian phase},\\
L_yT_E,
& \text{non-Abelian phase},
\end{cases}
\label{eq:kitaev_FNL_phase_diagnostic}
\end{equation}
where the $\sqrt{T_E}$ prefactor in the first line assumes a generic
quadratic minimum of the gapped entanglement band as $T_E\rightarrow0$.


\section{Quasi-particle picture}
We consider the dynamics on the cylinder~\eqref{eq:cylinder} following a quantum quench under the Hamiltonian~\eqref{eq:hopping}, setting the chemical potential $\mu=0$ for simplicity. The system is initialized in a translationally invariant, low-entanglement state with a fixed number of fermions, examples of which are discussed in~\cite{yamashika2024asymmetry,yamashika2024timeevolution}. Considering a shift-symmetric state with a fundamental cell defined by $\boldsymbol{\Delta}=(\Delta_x,\Delta_y,\dots)$, this can be expressed in the most general way as~\cite{bertini2018entanglement} 
\begin{equation}
\label{eq:symm_preserv_general}
\ket{\psi}= \prod_{\boldsymbol{j}=1}^{\boldsymbol{L}/\boldsymbol{\Delta}} \prod_{\lambda=1}^{\mathtt{n}}(\sum_{\boldsymbol{m}_\lambda=0}^{\boldsymbol{\Delta}-1}a^{(\lambda)}_{\boldsymbol{m}_\lambda} c_{\boldsymbol{\Delta} \boldsymbol{j}-\boldsymbol{m}_\lambda}^\dagger)\ket{0},
\end{equation}
where $a^{(\lambda)}_{\boldsymbol{m}_\lambda}$ are numerical coefficients, $\boldsymbol{\Delta}\boldsymbol{j}=(\Delta_xj_x, \Delta_yj_y,...)$ 
and we have used the compact notation
 \begin{equation}
\prod_{\boldsymbol{j}=1}^{\boldsymbol{L}/\boldsymbol{\Delta}} = \prod_{j_x=1}^{L_x/\Delta_x}\dots\prod_{j_d=1}^{L_d/\Delta_d},\hspace{0.5cm} \sum_{\boldsymbol{m}=0}^{\boldsymbol{\Delta}-1} = \sum_{m_x=0}^{\Delta_x-1} \dots\sum_{m_d=0}^{\Delta_d-1}
 \end{equation}

As a general feature of integrable out-of-equilibrium physics, post-quench dynamics in one-dimensional free-fermionic systems is accurately captured (for system sizes and times large enough compared to microscopic length scales) by the quasi-particle picture (QPP): upon the global quench, the initial state acts as a source of entangled quasi-particle pairs that are emitted locally and propagate ballistically in opposite directions with equal and opposite momenta, ultimately governing the spread of entanglement as well as the local relaxation of physical observables. This leads to results regarding the evolution of several entanglement-related quantities, including entanglement entropy, entanglement negativity, and symmetry-resolved entanglement measures \cite{coser2014entanglement,ParezBonsignori2021,murciano2022negativity,turkeshi2022,turkeshi2023,rottoli2024,Travaglino2025measurements}. The structure of the QPP implies that the  time dependence of the Rényi entropies admits a simple structure: since each pair of quasi-particles shared between $A$ and the complement carries an entanglement contribution $\tilde s_\alpha(\vartheta(k))=s_\alpha(2\vartheta( k)-1)$, where $\vartheta(k) = \expval*{c_{k}^\dag c_{ k}}$ is the momentum space occupation functions of Dirac fermions and $s_\alpha$ was defined in \eqref{eq:entropycontribution}, the spreading of correlations reduces to a counting of the number of shared pairs.
It is simple to realise that for each mode $k$, since quasi-particles propagate at a velocity $v_k = \partial_k \varepsilon_k$, such number is just $ \min(2t|v_k|,\ell_A)$, leading to 
\begin{equation}
S_\alpha(A) =  \int \frac{dk}{2\pi} \min(2t|v_k|,\ell_A)  \tilde s_\alpha(\vartheta( k))\,.
\end{equation}

In higher dimensional strip geometries, following quenches from states of the form \eqref{eq:symm_preserv_general}, Ref.~\cite{travaglino2024} demonstrated that the quasi-particle picture applies to the dynamics of the reduced density matrix itself through the entanglement Hamiltonian, which admits a simple expression in terms of the quasi-particle number operators.  
Since the entanglement Hamiltonian of particle number preserving free-fermionic states is related to the correlation matrix \eqref{eq:corrmat} through \cite{EHFF2}
\begin{equation}
    (k_A)_{\bm i \bm j} = \log \left(\frac{1-C_A}{C_A}\right)_{\bm i \bm j},
\end{equation}
where $K_A = \sum_i (k_A)_{\bm i \bm j}c_{\bm i}^\dag c_{\bm j}$, it is clear that any function of the correlation matrix inherently obeys a form of the quasi-particle picture.
In particular, observing that the eigenvalues $\xi_j$ of $C_A$ are related to the positive eigenvalues of $i\Gamma_A$ as in \eqref{eq:kitaev_lambda_xi}, the fermionic non-local magic can be expressed as
\begin{equation}
M_\alpha^{\FNL}(A)=\sum_j^{|A|} \tilde m_\alpha(\xi_j),
\label{eq:FNLformulacorrmat}
\end{equation}
where we have introduced the function 
\begin{equation}
    \tilde m_\alpha(\xi_j) = m_\alpha((2\xi_j-1)^2). \label{eq:tildem}
\end{equation}

The quasi-particle prediction for expressions of the form \eqref{eq:FNLformulacorrmat} arises in the regime of large $\ell$, in which the $x$ components of momenta can be assumed to take continuous values, while all other compactified directions can take arbitrary values (compatible with the symmetries of the initial states).
 A common feature of hydrodynamic treatments is that the slow dynamics determining the leading order behavior is provided by the transport of conserved quantities, as non-conserved modes decay exponentially, or are only responsible for small oscillations about the hydrodynamic value. In  free fermionic systems, such conserved modes are given by the full set of momentum space occupation functions of the Dirac fermions $\vartheta(\bm k)$. 
 Following the discussion of \cite{travaglino2024}, 
 the result for the non-local magic can therefore be constructed in complete analogy with the Rényi entropies. Because of the strip geometry, only the quasi-particle velocities in the $x$ direction matter for the counting, hence leading to a counting factor $(2t|v_x|,\ell)$; the contribution of each pair is then obtained by evaluating \eqref{eq:tildem} in the conserved occupation function. In two dimensions, this leads to the general results
\begin{equation}
\label{eq:qpp}
\begin{split}
&S_\alpha(A) =  \sum_{k_y}\int \frac{dk_x}{2\pi} \min(2t|v_x|,\ell)  \tilde s_\alpha(\vartheta(\bm k))  \\
 &   M_\alpha^{\FNL}(A) = \sum_{k_y}\int \frac{dk_x}{2\pi} \min(2t|v_x|,\ell) \tilde m_\alpha(\vartheta(\bm k)),
    \end{split}
\end{equation}
and the addition of further compactified dimensions is immediate. 
 Both expressions are valid as long as the transported quasi-particles are the fermions $c_{\bm x},c_{\bm x}^\dag$ appearing in the Hamiltonian. For some exotic initial states, the transported quasi-particles can be Bogoliubov transformations of the original fermions: this occurs, for example, in 1d for the tilted Néel state \cite{ares2023lack}, and in 2d for the crossed dimer state \cite{yamashika2024timeevolution}. In such cases, the result is simply modified by making use of the occupation functions of the Bogoliubov rotated fermionic operators and possibly restricting the domain of integration to account for a modified Brillouin zone.  

 The structure of the result implies that non-local magic and entanglement entropy are transported identically following a quench. This is a direct consequence of the general QPP formulation at the level of the reduced density matrix, which subsequently dictates the behavior of various entropic quantities. However, since the FNL and entropy contents of each pair is described by different functions $\tilde s_\alpha$ and $\tilde m_\alpha$, which are differently weighted across the entanglement spectrum, the two quantities can exhibit drastically different features along the evolution, as shown in figure \eqref{fig:qpp}.

\subsection{Numerical evaluations}
\begin{figure*}[ht]
    \centering
    \includegraphics[width=0.9\linewidth]{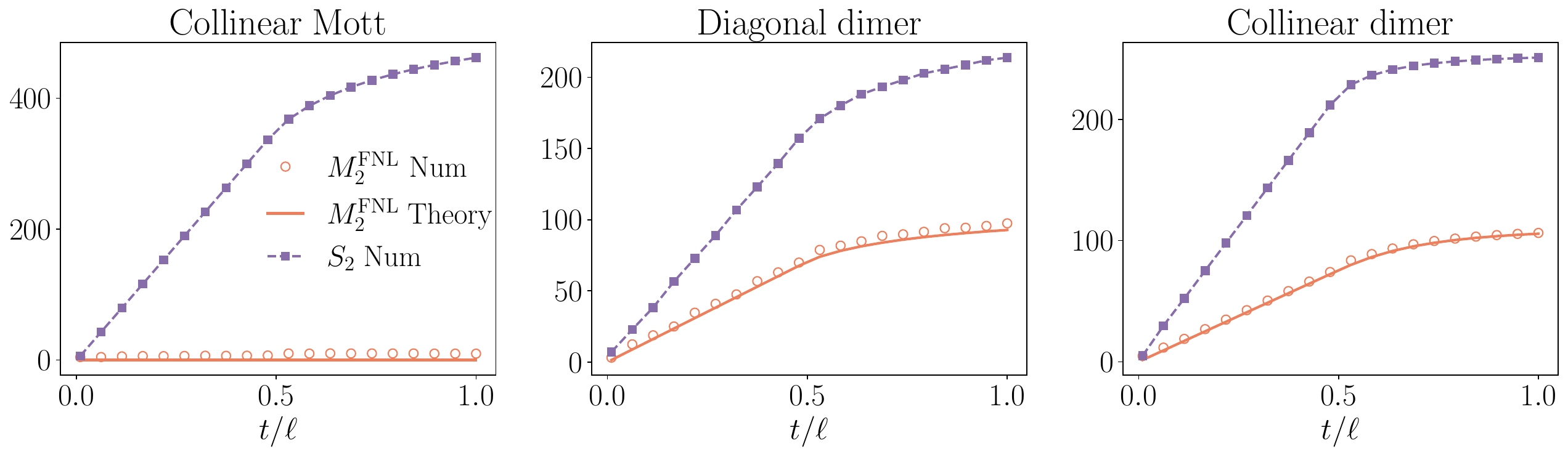}
    \caption{Quench dynamics of entanglemnt Rényi entropy and non-local magic for $\alpha=2$ in a quench from three different states, in a hopping hamiltonain with $L_y=10$ and $\ell=100$. The figure shows the perfect agreement with the exact numerical expression for the FNL \eqref{eq:FNLformula} and its quasi-particle version \eqref{eq:qpp}, and their comparison with the $\alpha=2$ Rènyi entropy, which shows clearly that the bound in \eqref{eq:M_S} is a loose bound. Interestingly, the figure makes clear that for collinear mott state, while the entanglement entropy grows extensively, the FNL remains close to zero at all times, highlighting that FNL truly plays a complementary role to entanglement, and can be small when entanglement is large. In general, this is reflecting the fact that the FNL is minimal when the Rényi entropies are maximal, namely for $\xi_i=1/2$ ($\lambda_i=0$), which correspond to the maximally mixed eigenvalues.}
    \label{fig:qpp}
\end{figure*}
Here we test the quasi-particle prediction \eqref{eq:qpp} against exact numerics obtained through correlation matrix techniques. Considering general shift-invariant states of the form \eqref{eq:symm_preserv_general}, the correlation matrix can often be evaluated exactly in closed form in terms of Bessel functions \cite{Eisler_2007,Peschel_2009}, allowing the evaluation of essentially arbitrary system sizes. The simplest kind of states admitting a dimensional reduction treatment are those which have a pure 1-site shift symmetry along the compactified direction, as in such cases the decomposition in \eqref{eq:widom_block}  is obvious. Two relevant examples, to which we will refer to as the collinear Mott and collinear Dimer states to match the nomenclature of \cite{yamashika2024timeevolution,travaglino2024}, are 
\begin{equation}
\begin{split} 
    \ket{\text{CM}} &= \prod_{x=1}^{L_x/2}\prod_{y=1}^{L_y} \frac{1}{\sqrt{2}}c^\dagger_{2x,y} \ket{0},\\
    \ket{\text{CD}} &= \prod_{x=1}^{L_x/2}\prod_{y=1}^{L_y} \frac{1}{\sqrt{2}}\left(c^\dagger_{2x,y} -c^\dagger_{2x+1,y}\right) \ket{0}.
\end{split}
\end{equation}
The quench dynamics of the corresponding correlation matrices can be simply evaluated, and leads to a perfectly decomposed structure at all times 
\begin{equation}
    C_{\boldsymbol{x},\boldsymbol{x}'}(t)= \delta_{y,y'}C_{x,x'}(t)
\end{equation}
where the 1D correlation matrices for the two states are 
\begin{equation}\label{eq:cd_correl}
\begin{split}
C^{\rm CM}_{x,x'}(t) &= \frac{1}{2}\left(\delta_{x,x'}-  e^{-i\pi/2 (x+x')}J_{x-x'}(2t)\right)\\
       C^{\rm CD}_{x,x'}(t) &= C^{\rm CD}_{\infty}-e^{-i\pi/2 (x+x')}\frac{i(x-x')}{4t}J_{x-x'}(2t)
\end{split}
\end{equation}
where $J_x(z)$ is the Bessel function of the first kind, and $C^{\rm CD}_{\infty}=\frac{1}{2}\left(\delta_{x,x'} - \frac{1}{2}(\delta_{x,x'+1}+\delta_{x,x'-1})\right)$. The structure of the correlation matrix immediately implies that the occupation functions are only dependent on  the x component of momentum,leading to a simplification of  the sum in $k_y$ appearing in \eqref{eq:qpp} to a simple multiplicative factor $L_y$.

A more interesting structure can be obtained by considering initial states with 2-site shift invariance along both directions. An example is the diagonal dimer, which exhibits a dimerization along the diagonal of the lattice,
\begin{equation}
\label{eq:DD}
    \ket{\text{DD}} = \prod_{x=1}^{L_x/2}\prod_{y=1}^{L_y} \frac{1}{\sqrt{2}}\left(c^\dagger_{2x,y} -c^\dagger_{2x+1,y+1}\right) \ket{0}.
\end{equation}
In this case the correlation matrix is consistent with a 2-site shift symmetry \cite{travaglino2024},
\begin{equation}
\begin{split}
      C_{\boldsymbol{x}, \boldsymbol{x'}}^{\rm DD}(t) &= C^{\rm DD}_{(\infty)} + \frac{\delta_{y+1,y'}}{4}i^{1-x-x'}J_{x-x'+1}(2t) \\ &+\frac{\delta_{y-1,y'}}{4}i^{1-x-x'}J_{x-x'-1}(2t),
\end{split}
\end{equation}
where the asymptotic value is now fixed to 
\begin{equation}
    C^{\rm DD}_{(\infty)}= \frac{1}{2}\delta_{\boldsymbol{x}, \boldsymbol{x'}} -\frac{1}{4} (\delta_{x',x+1}\delta_{y',y+1}+\delta_{x',x-1}\delta_{y',y-1}).
\end{equation}

\begin{figure}
    \centering
    \includegraphics[width=.7\linewidth]{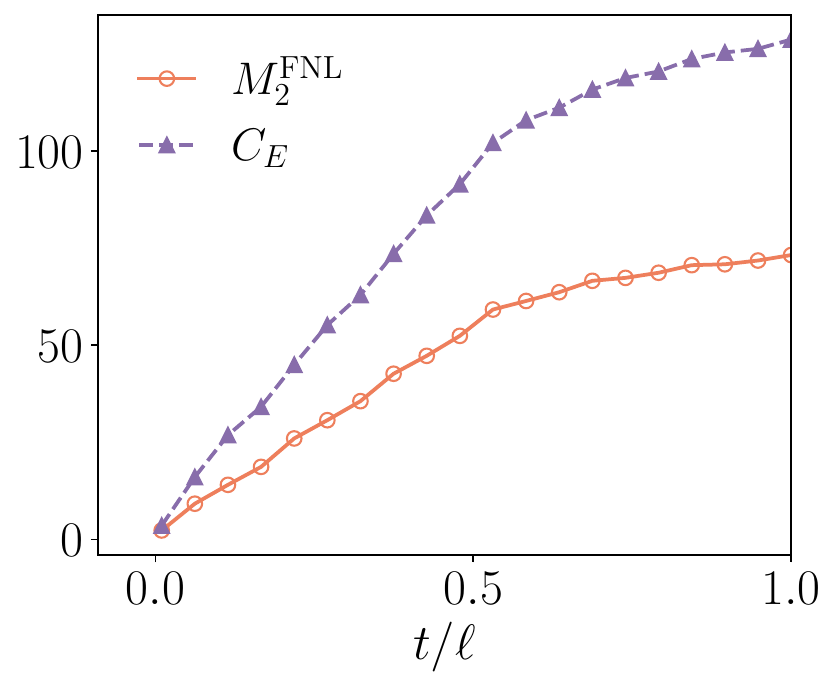}
    \caption{Comparison between FNL and capacity of entanglement in a quench from the diagonal dimer state \eqref{eq:DD}, which highlights that $C_E$ provides a closer bound to $M_2^{FNL}$ compared to $2S_2$.}
    \label{fig:qpp_ce}
\end{figure}

The dynamics of FNL following quenches from the three states are shown in figure \ref{fig:qpp}, demonstrating excellent agreement between the exact correlation matrix result and the quasiparticle prediction, and highlighting that the bound provided by the second Rényi entropy is generally highly overestimated. It is possible to see in figure \ref{fig:qpp_ce} that the capacity of entanglement bound provides a much tighter one. It is also interesting to observe that some quench scenarios, such as that originating from the collinear Mott state, exhibit extensive entanglement growth but a FNL which remains $O(1)$ at all times, highlighting how the two quantities can access complementary notions of complexity of a quantum state. 

\section{Random Gaussian circuits}
We consider a discrete-time protocol in which each time step consists of four successive layers of two-site fermionic Gaussian unitaries $U=U_e^{(x)} U_e^{(y)} U_o^{(x)} U_o^{(y)}$.
The superscript $r\in{x,y}$ specifies the spatial direction along which the gates act, while the subscripts $\mathrm{e}$ (even) and $\mathrm{o}$ (odd) distinguish the two alternating brickwork layers. Each layer is defined as $U_{\mathrm{e}/\mathrm{o}}^{(r)}=\bigotimes_{\langle i,j\rangle_{\mathrm{e}/\mathrm{o}}^{(r)}}
U_{ij}^{(r)}$, where $\langle i,j\rangle_{\mathrm{e}/\mathrm{o}}^{(r)}$ denotes the set of disjoint even or odd couples of nearest-neighbour bonds oriented along the direction $r$. The operator $U_{ij}^{(r)}$ is a random two-site fermionic Gaussian unitary acting nontrivially only on sites $i$ and $j$, and hence on the four Majorana operators $\gamma_{2i}$, $\gamma_{2i+1}$, $\gamma_{2j}$, and $\gamma_{2j+1}$. Since the gates belonging to the same layer act on disjoint pairs of sites, they commute and can be applied simultaneously.

\subsection{Diffusive behavior}
 We first study the dynamics of the average entanglement in two dimensions, $\overline{S_2}=-\mathbb{E}[\log \mathrm{tr}\rho_A^2]$, which we bound through the following insight. 
 
\begin{figure}
    \centering
    \includegraphics[width=0.9\linewidth]{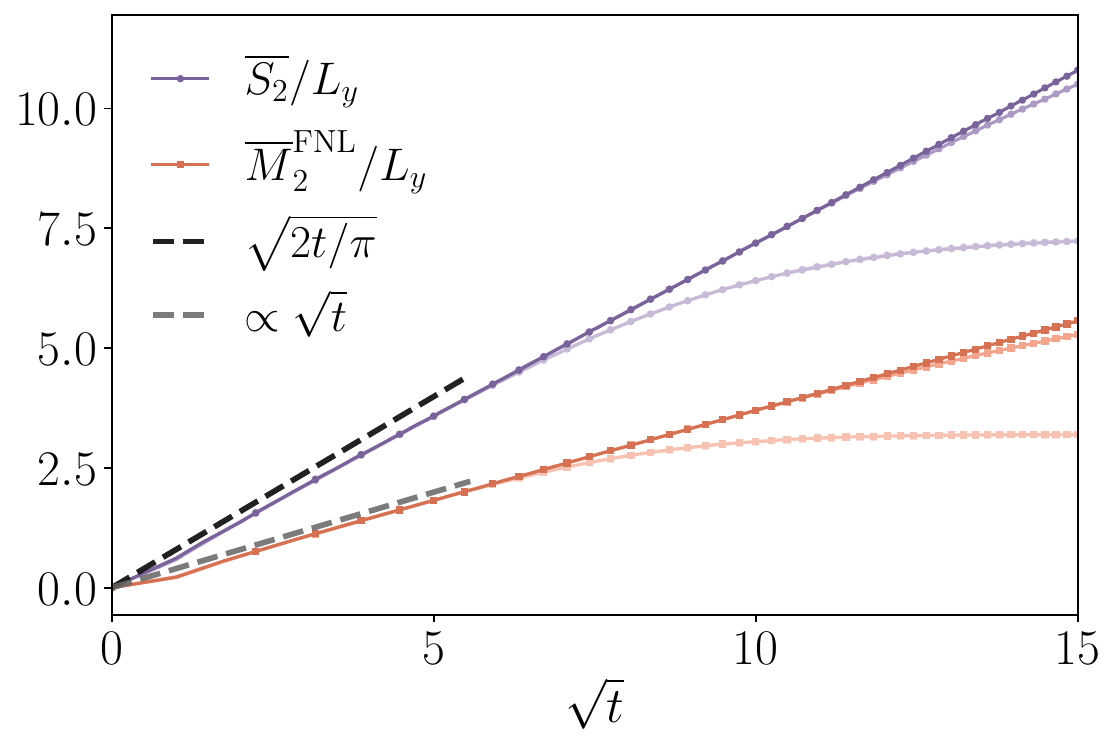}
    \caption{Numerical benchmarking of the dynamics of entanglement (purple) and fermionic non-local magic (orange) obtained by averaging over $n_{\rm sample} = 40$ up to depth $t=250$. Different shades represent different system sizes, $(L_x, L_y) = [(32,8), (32,32), (64,8), (128,8)]$, with the ones with the same $L_x$ overlying exactly. Both entanglement and FNL clearly manifest diffusive behavior (dashed lines) before reaching their maximum value at approximately $t = O(L_x^2)$.}
    \label{fig:circuit_dynamics}
\end{figure}

 Instead of considering the full Gaussian set, we take $U_{ij}^{(r)}$ to be a random Clifford Gaussian gate. The Clifford group, $\mathcal{C}_n$, is the normalizer of the Pauli group, meaning that the adjoint action of $C \in \mathcal{C}_n$ maps a Pauli string to another Pauli string, up to phases: $C^\dagger P C = P'$. Due to the Jordan-Wigner mapping, gates that are both Clifford and Gaussian act on Majorana operators as permutations, i.e. for $U\in \mathcal{CG}_n$ $U^\dagger\gamma_iU= \pm \gamma_{\sigma(i)}$ with $\sigma \in \mathcal{S}_{2n}$, the permutation group over $2n$ elements. 
In Ref.~\cite{sierant2026theorymatchgatecommutant}, Sierant et al. found that the Clifford Gaussian ensemble forms a 3-design with respect to the Gaussian ensemble, which implies that every averaged quantity over random Clifford Gaussian gates matches the same average over random Gaussian gates up to the third moment. Specifically, the annealed average 2-Rényi entropy
$\widetilde{S}_2^{\mathcal{CG}} = -\log[\mathbb{E}[\mathrm{tr}(\rho_A^2)]]$ over a random Clifford Gaussian circuit determines the same average $\widetilde{S}_2^{\mathcal{G}}$ in a Gaussian circuit. Moreover, Ref.~\cite{Swann_2025} showed that the latter corresponds, at leading order, to $\overline{S_2}$, thus implying
\begin{equation}
    \overline{S_2} \approx \widetilde{S}_2^{\mathcal{CG}} \leq \overline{S_2}^{\mathcal{CG}},
\end{equation}
due to Jensen's inequality.
Appendix~\ref{app:clifford_entanglement} presents the step-by-step procedure from which we obtain that for a vertical cut at $L_x/2$, in the scaling limit $L_x\rightarrow \infty$, 
\begin{equation}
    \overline{S_2}^{\mathcal{CG}}(t) \approx L_y\sqrt{2t/\pi}.
    \label{eq:diffusive_S}
\end{equation}
This result bounds the average dynamics in Gaussian circuits. The numerics in Fig.~\ref{fig:circuit_dynamics} show that
the two-dimensional random Gaussian circuit maintains the diffusive behavior of its one-dimensional version, differing by a constant factor from Eq.~\eqref{eq:diffusive_S}. 

Additionally, in Appendix~\ref{app:diffusive_gaussian_circuit}, we verified that the dynamics of a single Majorana mode is diffusive in any number of spatial dimensions. Hence, since the fermionic non-local magic is upper-bounded by the 2-Rényi entropy and directly depends on the spreading of Majorana modes, we infer that
\begin{equation} 
\overline{M}^{FNL}_2(t) \propto L_y\sqrt{t}.
\end{equation}
at $L_x/2$.
Numerical simulations, conducted at different system sizes, confirm this behavior (see Fig.~\ref{fig:circuit_dynamics}).

\section{Conclusion and outlook}

In this work, we investigated fermionic non-local magic in higher-dimensional free-fermion systems, studying its relation to the entanglement spectrum. 
For translationally invariant strip geometries, dimensional reduction expresses FNL as a sum over one-dimensional momentum sectors. A finite density of critical sectors then produces a multiplicative logarithmic violation of the boundary law, $M_\alpha^{\FNL}\sim L_y\log\ell$, analogous to Gioev--Klich--Widom scaling. For the square-lattice hopping model, we obtained an explicit asymptotic expression that relates this growth to the geometry of the Fermi surface.

We also established an upper bound on the second Rényi FNL in terms of the capacity of entanglement for free fermions, connecting once again non-local magic to fluctuations of the entanglement Hamiltonian~\cite{Cao_2025}. 
The entanglement-temperature deformation introduced here makes this link more explicit by resolving the contribution of low-energy entanglement modes. In the Gaussian matter sector of the Kitaev model, its low-temperature behavior distinguishes the gapped entanglement spectrum of the Abelian phase from the gapless branches of the non-Abelian phase.

Out of equilibrium, we extended the quasi-particle description to FNL following the quadratic quenches considered here. Entanglement and FNL share the same ballistic propagation mechanism while assigning different weights to the transported modes. This distinction allows extensive entanglement production to coexist with a vanishing leading hydrodynamic contribution to FNL, as illustrated by the collinear Mott quench. For random Gaussian circuits, our numerical results instead support diffusive growth before saturation, consistent with the diffusion of Majorana weight derived analytically. These results show how a common transport mechanism can govern quantum resources with different sensitivities to the entanglement spectrum.

Several questions remain open. Extending the equilibrium analysis to general subsystem geometries would clarify how boundary and Fermi-surface geometry jointly determine the leading FNL contribution.
Beyond pure states, the existence of magic carried entirely by correlations in separable states~\cite{wei2026entirelynonlocalquantummagic} motivates investigating the relation between this distinct notion of non-local magic and suitable mixed-state extensions of the present framework~\cite{esposito2026stabilizerentropytrustworthymixed}.
Finally, since phononic platforms with a fermion-filling analogue already allow the reconstruction of free-fermion correlation matrices and entanglement spectra~\cite{Lin_2024}; one can directly test the FNL spectral function against these reconstructed eigenvalues, probing the boundary-law violation for non-local magic.

\section*{Acknowledgments}
We thank Xhek Turkeshi and Alioscia Hamma for discussions and related works on similar topics.

D.I. acknowledges support of Research Grants in Germany, 2026 (57812126) from the German Academic Exchange Service (DAAD). R.T. acknowledges support by the ERC-AdG grant MOSE No. 101199196.
B.M. acknowledges support from DFG Emmy Noether Programme proposal "Digital Quantum Matter Out-of-Equilibrium" No. 560726973, DFG under Germany's Excellence Strategy – Cluster of Excellence Matter and Light for Quantum Computing (ML4Q) EXC 2004/2 – 390534769, and DFG Collaborative Research Center (CRC) 183 Project No. 277101999 - project B01.

\textit{Code and Data Availability}.– The code and the data for
our simulations will be publicly shared at publication.

\bibliographystyle{apsrev4-2}
\bibliography{ref}

\clearpage

\appendix
\onecolumngrid

\section{Scaling behavior $M_2^{\FNL}$ at criticality for the two dimensional hopping model}
\label{eq:scalingMFNL}

We now derive the asymptotic scaling of the fermionic non-local magic at criticality. The derivation follows the Fisher--Hartwig approach introduced for the entanglement entropy of free fermions in Ref.~\cite{Jin_2004,Murciano_2020}, with the entropic function replaced by the single-mode contribution to the fermionic non-local magic,
\begin{equation}
m_2(x)
=
-\ln(1-x+x^2).
\label{eq:log_m2}
\end{equation}
Let us start with the one-dimensional system, namely the one-dimensional tight-binding model along $y$ in the two-dimensional system at hand, and define
\begin{equation}
    D_\ell(z)=\det\left(z\mathbb{I}- \Gamma_A \right)= \prod_{j=1}^\ell (z-\lambda_j)\,,\quad \partial_z \ln D_\ell(z) =\sum_{j=1}^\ell \frac{1}{z-\lambda_j}\,.
\end{equation}
where $\lambda_j\in(-1,1)$ are the eigenvalues entering the fermionic non-local magic.  
\begin{figure}[h!]
    \centering
    \includegraphics[width=0.7\linewidth]{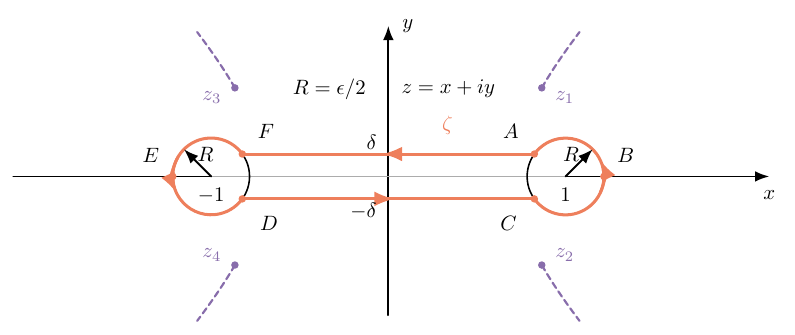}
    \caption{
Contour $\zeta$, adapted from Ref.~\cite{Jin_2004}. 
The zeros of $D_{\ell}(z)$ lie on the interval $(-1,1)$, which becomes the branch cut of $\partial_z\ln D_{\ell}(z)$ in the limit $\ell\to\infty$. 
The function $m_2(z^2)=-\ln(1-z^2+z^4)$ has four branch points,
$z_1=e^{i\pi/6}$, $z_2=e^{-i\pi/6}$, $z_3=e^{5i\pi/6}$, and $z_4=e^{-5i\pi/6}$,
from which the corresponding branch cuts extend into the complex plane. 
The contour $\zeta$ encloses the interval $(-1,1)$ while avoiding these branch cuts. 
The arrows indicate the orientation of the contour integration, while $R=\epsilon/2$ denotes the radius of the circular arcs centred at $z=\pm1$.
}
    \label{fig:complexintegral}
\end{figure}
Then, using Cauchy's theorem, we have
\begin{equation}
     \lim_{\epsilon \to 0^+} \lim_{\delta \to 0^+} \frac{1}{2 \pi i }\oint_{\zeta} m_2(z^2) \,\partial_z \ln D_\ell(z) \, dz = \lim_{\epsilon \to 0^+} \lim_{\delta \to 0^+} \sum_{j=1}^\ell  \frac{1}{2 \pi i }\oint_{\zeta} \frac{m_2(z^2)}{z-\lambda_j} \, dz= \sum_{j=1}^{\ell} m_2(\lambda_j^2)= M_2^{\FNL}(A)\,,
\end{equation}
where $\zeta$ is the contour represented in Fig.~\ref{fig:complexintegral} and depends $\epsilon$ and $\delta$.
The logarithm~\eqref{eq:log_m2} introduces four branch points determined by $1-z^2+z^4=0$ that lie away from $[-1,1]$, i.e.
$z_1=e^{i\pi/6},
\;
z_2=e^{-i\pi/6},
\;
z_3=e^{5i\pi/6},
\;
z_4=e^{-5i\pi/6}.$

We can therefore choose the contour $\zeta(\epsilon,\delta)$ so that it encloses the complete interval $[-1,1]$, and consequently all the eigenvalues $\lambda_j$, while leaving the branch points and branch cuts of $m_2(z^2)$ outside the contour.
At large $\ell$, the Fisher--Hartwig expansion of the Toeplitz determinant gives
\begin{equation}
\frac{d \ln D_\ell(z)}{dz}
=
\left(
\frac{1-k_F/\pi}{1+z}
-
\frac{k_F/\pi}{1-z}
\right)\ell
-
\frac{4}{i\pi}
\frac{\xi(z)}{(1+z)(1-z)}
\left(\ln(2 \ell|\sin k_F|)+1+\gamma_E+\Upsilon(z)
\right),
\end{equation}
with $\xi(z)=\frac{1}{2 \pi i} \ln \frac{1+z}{1-z}$ possessing a branch cut along $(-1,1)$, $\Upsilon(z)= \sum_{n=1}^\infty \frac{n^{-1} \xi^2(z)}{n^2-\xi^2(z)}$ and $\gamma_E$ the Euler constant.

We can now separate the extensive contribution from the Fisher--Hartwig correction,
\begin{equation}
M_2^{\FNL}(A)
=
M_{2,\mathrm{vol}}^{\FNL}
+
M_{2,\mathrm{FH}}^{\FNL}
+
o(1).
\end{equation}
The first term vanishes due to the symmetry of $m_2$
\begin{equation}
    M_{2,\mathrm{vol}}^{\FNL} = \lim_{\epsilon \to 0^+} \lim_{\delta \to 0^+} \frac{1}{2 \pi i }\oint_{\zeta(\epsilon,\delta)} m_2(z^2) \left(
\frac{1-k_F/\pi}{1+z}
-
\frac{k_F/\pi}{1-z}
\right)\ell \, dz =0 \,.
\end{equation}

We are thus left with the Fisher--Hartwig contribution,
\begin{equation}
M_{2,\mathrm{FH}}^{\FNL}
=
\frac{2}{\pi^2}
\lim_{\epsilon\to0^+}
\lim_{\delta\to0^+} \oint_{\zeta(\epsilon,\delta)} dz\, \frac{m_2(z^2)\xi(z)}{1-z^2} \left(\ln(2 \ell|\sin k_F|)+1+\gamma_E+\Upsilon(z)
\right) .
\label{eq:FNL_FH_contour}
\end{equation}
The contributions along the small circular arcs
centred at $z=\pm1$ vanish as their radius $R=\epsilon/2$ tends to zero.
Let $\overline{CBA} = C_+$ denote the arc surrounding $z=1$ and set $z=1+w$ with $|w|=R$.
We have that $m_2(z^2) =-2w+O(w^2),$
while $1-z^2=-2w+O(w^2)$, then it follows that $\frac{m_2(z^2)}{1-z^2}=1+O(w)$. Additionally, we also have $\xi(1+w)=-\frac{1}{2\pi i}\ln w+O(1)$.
Hence, close to $z=1$,
\begin{equation}
\frac{m_2(z^2)\xi(z)}{1-z^2}
=
O(\ln w).
\end{equation}
Parametrizing the circular arc as $w=Re^{i\theta}$,
its contribution to the term proportional to $\ln\ell$ is
\begin{equation}
\left|
\int_{C_+}
dz\,
\frac{m_2(z^2)\xi(z)}{1-z^2}
\right|
=
O\left(R|\ln R|\right) \xrightarrow{R\to 0} 0. 
\end{equation}
The same argument applies to the circular arc $C_-$ around $z=-1$.
We are now left to evaluate the upper end lower segments. Along them we have that $\xi(x\pm i\delta) =  \frac{1}{2i\pi} \left(\ln\frac{1+x}{1-x}\mp i (\pi + \delta)\right) = -i W(x)\mp (1/2+\delta) $, where $W(x) = \frac{1}{2\pi}\ln \frac{1+x}{1-x}$. Hence, the remaining integral becomes 
\begin{equation}
\begin{aligned}
M_{2,\rm FH}^{\rm FNL}
=& \frac{2}{\pi^2}
\int_{-1}^{1} dx\,
\frac{m_2(x^2)}{1-x^2}
\left(\ln (2\ell) + \ln\left(\left|\sin k_F\right|\right) + 1+\gamma_E\right)
\\
&\quad
+ \sum_{n=1}^{\infty}
\frac{2n^{-1}}{\pi^2}
\int_{-1}^{1} dx\,
\frac{m_2(x^2)}{1-x^2}
\left(
\frac{\left(\frac{1}{2}+iW(x)\right)^3}{n^2-\left(\frac{1}{2}+iW(x)\right)^2}
+\frac{
\left(\frac{1}{2}-iW(x)\right)^3}{n^2-\left(\frac{1}{2}-iW(x)\right)^2}\right).
\end{aligned}
\end{equation}
where we have fully expanded the definition of $\Upsilon(z)$. The first integral correspond to double the constant $\beta_2^{\rm FNL} = \frac{2}{\pi^2} \int_{0}^{1} dx\,
\frac{m_2(x^2)}{1-x^2}$ defined in Ref.~\cite{iannotti2026nonlocalmagicresourcesfermionic} while the second integral, which we call $\mathcal{J}_2$, is solved in the next section. With these substitutions, we obtain the one-dimensional FNL magic at criticality for the Hopping model, which, after summing over the critical modes along $y$, gives us the two-dimensional result, i.e. Eq.~\eqref{eq:FNL_square_lattice_scaling} of the Main Text. 
Explicitly, the Fermi momentum of the effective chain associated
with the transverse momentum $k=2\pi n/L_y$ satisfies
\begin{equation}
    \sin k_F
    =
    \sqrt{
        1-\left[
            \mu-\cos\left(\frac{2\pi n}{L_y}\right)
        \right]^2
    },
\end{equation}
where only modes fulfilling
$\left|\mu-\cos(2\pi n/L_y)\right|<1$
contribute to the critical scaling, with 
\begin{equation}
\begin{split}
     n \in \Omega_n &=\Big[0, \frac{\arccos(\mu-1) L_y}{2 \pi} \Big[ \;\cup\Big] L_y \left(1-\frac{\arccos(\mu-1) }{2 \pi}\right), L_y-1\Big]\,.
\end{split}
\end{equation}

\section{Constant term of $M_2^{\FNL}$ at criticality for one-dimensional Ising chain}
\label{App:const_Ising}
Using the results of Ref.~\cite{Jin_2004} we can determine the constant term evaluated numerically in 
Ref.~\cite{iannotti2026nonlocalmagicresourcesfermionic} for the transverse-field Ising chain at criticality. 
For the lattice normalization used in the main text,
\begin{equation}
    M_2^{\FNL}(\ell) \simeq \beta_{\FNL} \ln \ell + b ,
\end{equation}
where
\begin{equation}
    \beta_{\FNL}
    =
    \frac{2}{\pi^2}
    \int_0^1 d\lambda\,
    \frac{m_2(\lambda^2)}{1-\lambda^2}
    =
    \frac{\pi^2-\ln^2(7-4\sqrt{3})}{\pi^2\ln 16}.
\end{equation}
The Fisher--Hartwig constant is obtained by replacing the entropy kernel in Ref.~\cite{Jin_2004}
with the FNL kernel \(m_2(x)=-\log_2(1-x+x^2)\). This gives
\begin{equation}
    b =
    \beta_{\FNL}\ln 4
    -
    \frac{1}{2\pi^2}
    \int_{-1}^{1} d\lambda\,
    \frac{m_2(\lambda^2)}{1-\lambda^2}
    \left[
    \psi\!\left(\frac{1}{2}+iW(\lambda)\right)
    +
    \psi\!\left(\frac{1}{2}-iW(\lambda)\right)
    \right] \simeq 0.27958869083.
    \label{eq:b_fnl_ising}
\end{equation}
with $W(x)=\frac{1}{2\pi}\ln\frac{1+x}{1-x}$ and diagamma function $\psi(z)=\frac{d}{dz}\ln\Gamma(z)$.
Let us denote the integral appearing in Eq.~\eqref{eq:b_fnl_ising} by
\begin{equation}
    \mathcal J_2 =
    \int_{-1}^{1} d\lambda\,
    \frac{m_2(\lambda^2)}{1-\lambda^2}
    \left[
    \psi\!\left(\frac{1}{2}+iW(\lambda)\right)
    +
    \psi\!\left(\frac{1}{2}-iW(\lambda)\right)
    \right].
\end{equation}
Following Appendix A of Ref.~\cite{Jin_2004}, we introduce $ \lambda=\tanh(\pi w)$ and $W(\lambda)=w$.
Since the integrand is even, this gives
\begin{align}
    \mathcal J_2
    &=
    2\pi
    \int_0^\infty dw\,
    m_2\!\left(\tanh^2(\pi w)\right)
    \left[
    \psi\!\left(\frac{1}{2}+iw\right)
    +
    \psi\!\left(\frac{1}{2}-iw\right)
    \right].
    \label{eq:I2_w_form}
\end{align}
Using the same Gamma-function representation as in Ref.~\cite{Jin_2004}, this expression can be rewritten as
\begin{equation}
    \mathcal J_2
    =
    2\pi^2
    \int_0^\infty \frac{dt}{t}
    \left[
    \beta_{\FNL}e^{-t}
    -
    \frac{2}{\ln 2}\,
    \frac{
    \cosh(t/4)\cos\!\left(\frac{rt}{2\pi}\right)-1
    }{
    \sinh^2(t/2)
    }
    \right].
    \label{eq:I2_t_form}
\end{equation}
which is a standard variation of the integral known as Barnes G-function identity~\cite{Jin_2004}, with $r=\ln(2+\sqrt{3})$.
Generalization to arbitrary $\alpha$ for $M_\alpha^{\FNL}$ involves  substituiting $m_{\alpha}(x)$ to $m_2(x)$.

\section{Comparison between Gaussian unitary minimization and full minimization of non-local magic}
\label{ap:Comparison}

\begin{figure}[h!]
    \centering
    \includegraphics[width=0.8\linewidth]{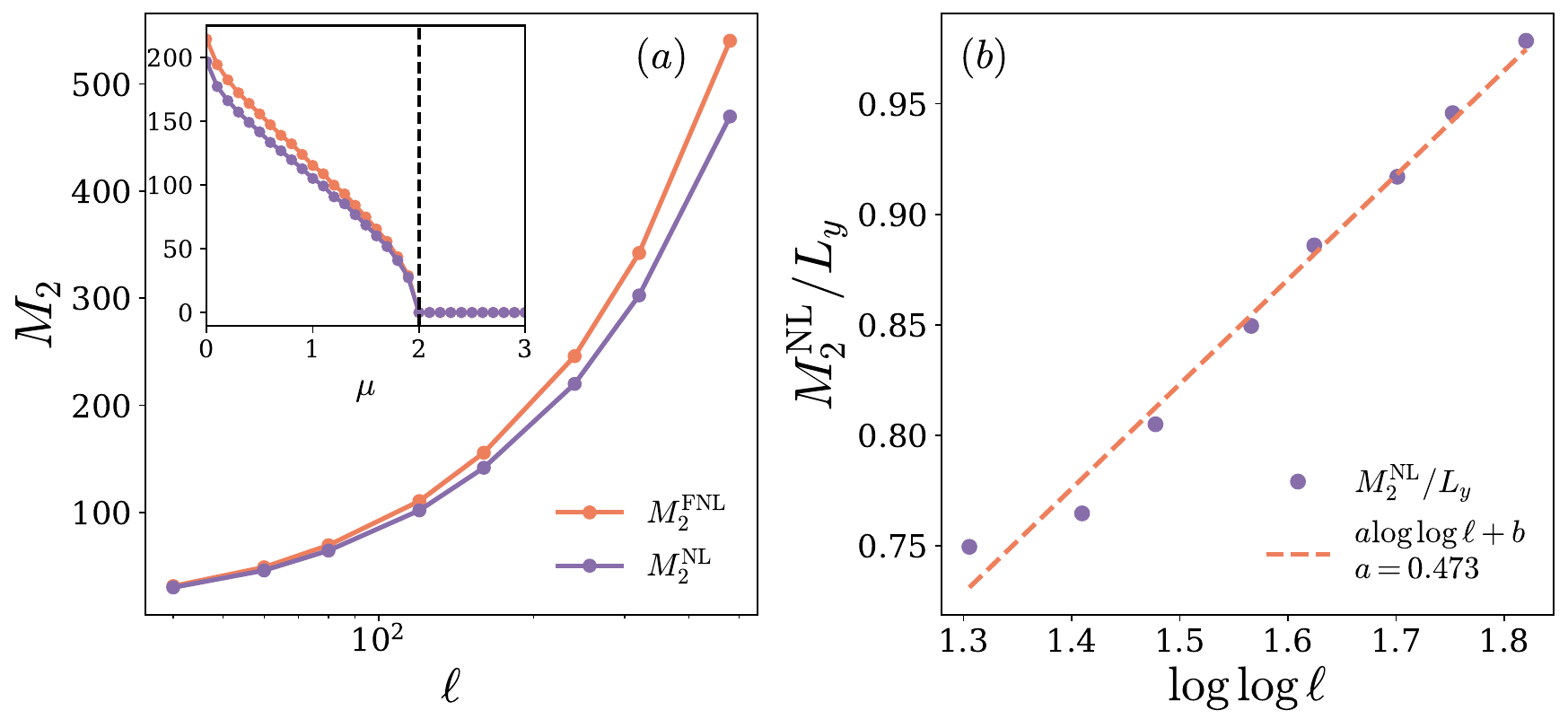}
    \caption{($a$) Comparison between Eq.~\eqref{eq:FNL_square_lattice_scaling} and the numerical values obtained from Eq.~\eqref{eq:xor}. The strips considered have dimension $L_y = \ell = \{40, 60, 80, 120, 160, 240, 320, 480\}$ with $\mu = 0.5$. In the inset, we show the same comparison of the FNL with fixed $L_y = \ell = 160$ and $\mu \in [0,3]$ varying inside and outside the critical regime (black dashed line). ($b$) Non-local magic in Eq.~\eqref{eq:xor} $L_y \log \log \ell$ behavior.}
    \label{fig:comparison}
\end{figure}

In the main text we defined the fermionic non-local magic by restricting
the minimization in Eq.~\eqref{eq:defNLmagic} to local Gaussian unitaries. Since the Gaussian unitary group is a subgroup of the full unitary group, one immediately has
\begin{equation}
M_2^{\NL}(A)\leq M_2^{\FNL}(A).
\label{eq:NL_FNL_inequality}
\end{equation}
For the second stabilizer Rényi entropy, a useful representation of the unrestricted problem is obtained directly from the Schmidt spectrum~\cite{Cao_2025,
sierant2026spectralgeometrynonlocalstabilizer,sierant2026exactquantificationnonlocalmagic,qian2025quantumnonlocalnonstabilizerness}.
Let $\{\pm\lambda_j\}$ denote the eigenvalues of $i\Gamma_A$, then the Gaussian Schmidt decomposition associates with them the many-body Schmidt eigenvalues
\begin{equation}
\Lambda_{\boldsymbol n}
=
\prod_{j=1}^{|A|}
\left(\frac{1+\lambda_j}{2}\right)^{1-n_j}
\left(\frac{1-\lambda_j}{2}\right)^{n_j},
\qquad
n_j\in\{0,1\}.
\label{eq:many_body_schmidt_spectrum}
\end{equation}
Unlike Eq.~\eqref{eq:FNLformula}, which is additive over the single-particle entanglement modes, the unrestricted local-unitary problem is sensitive to the organization of the full many-body entanglement spectrum.

To make this connection explicit, let
$\{\Lambda_i\}_{i=0}^{D-1}$ denote the many-body spectrum in
Eq.~\eqref{eq:many_body_schmidt_spectrum}, ordered such that
$\Lambda_i\geq\Lambda_j$ for $i<j$, with $D=2^{|A|}$.
Following Ref.~\cite{Cao_2025}, one can associate with this spectrum the
Schmidt-gauge state
\begin{equation}
|\psi_{\rm Sch}\rangle_{AB} =\sum_{i=0}^{D-1} \sqrt{\Lambda_i}\, |s_i\rangle_A |s_i\rangle_B,
\label{eq:schmidt_gauge_state}
\end{equation}
where $\{|s_i\rangle\}$ is a common eigenbasis of a stabilizer group.
For the second stabilizer R\'enyi entropy, this state provides an upper
bound to the unrestricted local-unitary minimum~\cite{Cao_2025}.
It is conjectured that this Schmidt-gauge representative actually
attains the minimum, while an analogous statement has been proved for
different measures of magic~\cite{sierant2026exactquantificationnonlocalmagic,
viscardi2026nonlocalmagicclosedformsolution}.

The second stabilizer R\'enyi entropy of
Eq.~\eqref{eq:schmidt_gauge_state} admits the exact spectral expression
\begin{equation}
\mathcal{M}_2(\{\Lambda_i\})
=-\log\Bigg[
\sum_{i_1,i_2,i_3,i_4=0}^{D-1}
\sqrt{
\Lambda_{i_1}
\Lambda_{i_2}
\Lambda_{i_3}
\Lambda_{i_4}
\Lambda_{i_1\oplus i_2\oplus i_3}
\Lambda_{i_1\oplus i_2\oplus i_4}
\Lambda_{i_1\oplus i_3\oplus i_4}
\Lambda_{i_2\oplus i_3\oplus i_4}
}
\Bigg],
\label{eq:xor}
\end{equation}
where $\oplus$ denotes the bitwise XOR operation. The value of
Eq.~\eqref{eq:xor} depends on the assignment of the Schmidt eigenvalues
to the stabilizer basis, and is minimized when the eigenvalues are
ordered in decreasing order~\cite{Cao_2025}. Therefore,
\begin{equation}
M_2^{\NL}(A)
\leq
\mathcal{M}_2(\{\Lambda_i\}),
\end{equation}
with equality conjectured for the second stabilizer Rényi entropy.
In Fig~\ref{fig:comparison} we show the $M_2^{\FNL}$ and $M_2^{\NL}$ for the nearest-neighbour hopping Hamiltonian~\eqref{eq:hopping}, assuming the conjecture holds.

\section{Scaling of $M_2^{\FNL}$ for a lowest entanglement band with a quadratic minimum}
\label{app:quadratic-minimum}

At low entanglement temperature, the contribution of a mode with entanglement energy $\varepsilon$ is exponentially suppressed. Indeed, using $\lambda(T_E)=\tanh[\varepsilon/(2T_E)]$, one finds
\begin{equation}
m_2\!\left(\lambda^2(T_E)\right)
=
4\ee^{-\varepsilon/T_E}
+
O\!\left(\ee^{-2\varepsilon/T_E}\right).
\end{equation}

For large $L_y$, the momentum sum can be replaced by $(L_y/2\pi)\int dk_y$. If the lowest entanglement band has a nondegenerate quadratic minimum $\varepsilon(k_y)=\Delta_E+a(k_y-k_0)^2+\cdots$, with $a>0$, only a momentum window of width $|k_y-k_0|\sim\sqrt{T_E/a}$ contributes. Extending the resulting Gaussian integral to the real line gives
\begin{equation}
M_2^{\FNL}(T_E)
\simeq
\frac{2L_y}{\pi}\,
\ee^{-\Delta_E/T_E}
\int_{-\infty}^{\infty}dq\,
\ee^{-aq^2/T_E}
=
\frac{2L_y}{\sqrt{\pi a}}\,
\sqrt{T_E}\,
\ee^{-\Delta_E/T_E}.
\label{eq:FNL_abelian_asymptotic}
\end{equation}

Therefore, up to a coefficient fixed by the curvature of the entanglement band,
$M_2^{\FNL}(T_E)\sim L_y\sqrt{T_E}\,\ee^{-\Delta_E/T_E}$. The exponential factor reflects the entanglement gap, while the factor $\sqrt{T_E}$ comes from the shrinking momentum window around the quadratic minimum. This asymptotic form assumes the continuum-momentum limit; at fixed finite $L_y$ and sufficiently small $T_E$, only the lowest discrete mode contributes.

\section{Diffusive behavior of a single Majorana mode in fermionic Gaussian brickwork circuits}
\label{app:diffusive_gaussian_circuit}

\begin{figure}
    \centering
    \includegraphics[width=0.8\linewidth]{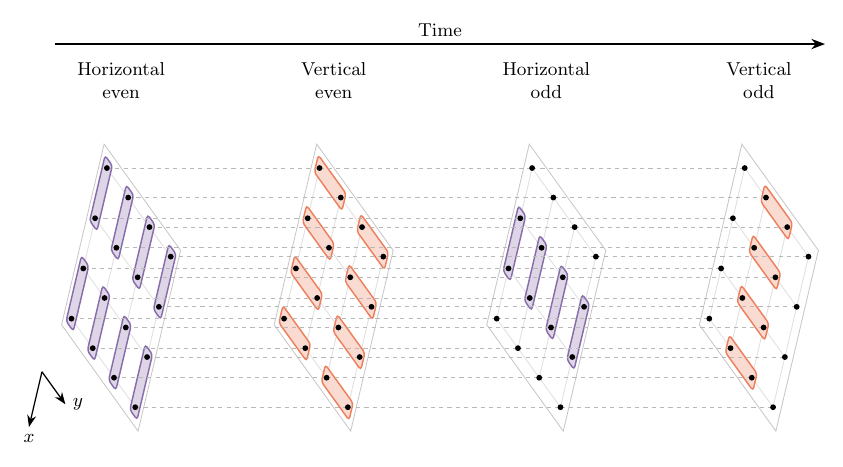}
   \caption{ One time step of a two-dimensional random Gaussian brickwork circuit on a lattice with open boundary conditions in both spatial directions. Time runs from left to right through four sublayers: horizontal even, vertical even, horizontal odd, and vertical odd. Violet and peach rectangles represent independently sampled two-site Gaussian gates acting along $x$ and $y$, respectively.
}
    \label{fig:placeholder}
\end{figure}

We first consider a one-dimensional periodic chain of an even number $L$ of
complex fermions. For the sake of easier notation in this section, we rescale our index system such that $x$ starts from $0$ and ends at $L-1$.
The odd and even brickwork layers are defined by
\begin{equation}
    U_e
    =
    \bigotimes_{i=0}^{L/2-1}
    U_{2i,2i+1},
    \qquad
    U_o
    =
    \bigotimes_{i=1}^{L/2-1}
    U_{2i-1,2i},
    \label{eq:brickwork_layers}
\end{equation}
where we considered open boundary conditions.

At every time step, Majorana operators evolve following the order 
\begin{equation}
    \gamma_a\left(t+\frac12\right)
    =
    U_e^\dagger\gamma_a(t)U_e, \quad
    \gamma_a(t+1)
    =
    U_o^\dagger
    \gamma_a\left(t+\frac12\right)
    U_o.
    \label{eq:heisenberg_half_steps}
\end{equation}
specifically, a two-site fermionic Gaussian unitary $U_{i,j}$ acts linearly on four Majoranas $\left(\gamma_{2i},\gamma_{2i+1}, \gamma_{2j},\gamma_{2j+1}\right)$,
\begin{equation}
    U_{i,j}^\dagger\gamma_kU_{i,j}
    =
    \sum_{\ell\in S_{i,j}}
    \Omega_{k\ell}\gamma_\ell,
    \qquad
    S_{i,j}
    =
    \{2i,2i+1,2j,2j+1\},
    \label{eq:local_gaussian_action}
\end{equation}
where $\Omega\in O(4)$. We label the rows and columns of the local
$4\times4$ block by the corresponding global Majorana indices.
For $k\notin S_{i,j}$, one has $ U_{i,j}^\dagger\gamma_kU_{i,j}=\gamma_k$.

Consider the two Majoranas $\gamma_{2j+r}$ on site $j$, with $r=0,1$.
For the odd layer,
\begin{equation}
    U_o^\dagger\gamma_{2j+r}U_o
    =
    \begin{cases}
    U_{j-1,j}^\dagger
    \gamma_{2j+r}
    U_{j-1,j},
    & j\ \text{even},
    \\[2mm]
    U_{j,j+1}^\dagger
    \gamma_{2j+r}
    U_{j,j+1},
    & j\ \text{odd}.
    \end{cases}= 
    \begin{cases}
    \displaystyle
    \sum_{\ell=0}^{3}
    \Omega_{2j+r,\,2j-2+\ell} \,
    \gamma_{2j-2+\ell},
    & j\ \text{even},
    \\[4mm]
    \displaystyle
    \sum_{\ell=0}^{3}
    \Omega_{2j+r,\,2j+\ell} \,
    \gamma_{2j+\ell},
    & j\ \text{odd}.
    \end{cases}
    \label{eq:even_layer_cases}
\end{equation}
Introduce $\mu_o\in\{0,1\}$ by $\mu_o =  \begin{cases}    0, & j\ \text{even},  \\  1, & j\ \text{odd}\end{cases}$ with which we condense Eq.~\eqref{eq:even_layer_cases} as
\begin{equation}
\begin{aligned}
    &
    U_{j-1+\mu_o,\,j+\mu_o}^\dagger
    \gamma_{2j+r}
    U_{j-1+\mu_o,\,j+\mu_o}
    =
    \sum_{\ell=0}^{3}
    \Omega_{2j+r,\,2j-2+2\mu_o+\ell}
    \gamma_{2j-2+2\mu_o+\ell},
    \qquad
    r=0,1.
\end{aligned}
\label{eq:even_layer_condensed}
\end{equation}

Analogously, introduce $\mu_e\in\{0,1\}$ by
    $\mu_e
    =
    \begin{cases}
        0, & j\ \text{even},
        \\
        1, & j\ \text{odd}.
    \end{cases}$, which give, for the odd layer,
\begin{equation}
\begin{aligned}
    &
    U_{j-\mu_e,\,j+1-\mu_e}^\dagger
    \gamma_{2j+r}
    U_{j-\mu_e,\,j+1-\mu_e}
    =
    \sum_{\ell=0}^{3}
    \Omega_{2j+r,\,2j-2\mu_e+\ell}
    \gamma_{2j-2\mu_e+\ell},
    \qquad
    r=0,1.
\end{aligned}
\label{eq:odd_layer_condensed}
\end{equation}
For later convenience, we also introduce the layer-dependent displacement 
$m_\alpha
    =
    \begin{cases}
        -2+2\mu_o,
        & \alpha=o,
        \\
        -2\mu_e,
        & \alpha=e.
    \end{cases}$

Both transformations can then be written in terms of the four output Majoranas $(\gamma_{2j+m_\alpha},\gamma_{2j+m_\alpha+1},\gamma_{2j+m_\alpha+2},\gamma_{2j+m_\alpha+3} )$.

\subsubsection{Evolution of a single Majorana mode}

Let $\gamma_a$ be evolved by the circuit up to time $t$. We write
    $\gamma_a(t)
    =
    U^\dagger(t)
    \gamma_a
     U(t)
    =
    \sum_{b=0}^{2L-1}
    O_{ab}(t)\gamma_b,$
where $O(t)\in O(2L)$ contains the complete evolution generated by all previous layers.
Let $O^e$ denote the global orthogonal matrix induced by the even layer. Using linearity,
\begin{equation}
\begin{aligned}
    \gamma_a\left(t+\frac12\right)
    &=
    U_e^\dagger \gamma_a(t) U_e=
    U_e^\dagger
    \left[
    \sum_b O_{ab}(t)\gamma_b
    \right]
    U_e=
    \sum_b
    O_{ab}(t)
    U_o^\dagger\gamma_bU_o=
    \sum_{b,c}
    O_{ab}(t)
    O^e_{bc}
    \gamma_c=
    \sum_c
    O_{ac}\left(t+\frac12\right)
    \gamma_c.
\end{aligned}
\label{eq:half_step_global_evolution}
\end{equation}
Therefore $O\left(t+\frac12\right)=O(t)O^e$ and  similarly, for the odd layer, $O(t+1)=O\left(t+\frac12\right)O^o$.
Since $O$ is orthogonal, i.e. $\sum_b O_{ab}O_{cb}=\delta_{ac}$, we have $\sum_b O_{ab}^2=1$.
The same property holds for every local matrix $\Omega$,
$\sum_{\ell=0}^{3}\Omega_{k,\,2j+m_\alpha+\ell}^{\,2}=1.$
After acting with one local gate in layer $\alpha\in\{e,o\}$, define
\begin{equation}
    \bm w_{j,r}^{(\alpha)}
    =
    \begin{pmatrix}
    \displaystyle
    \sum_{\ell=0}^{1}
    \Omega_{2j+r,\,2j+m_\alpha+\ell}^{\,2}
    \\[4mm]
    \displaystyle
    \sum_{\ell=2}^{3}
    \Omega_{2j+r,\,2j+m_\alpha+\ell}^{\,2}
    \end{pmatrix},
    \qquad
    r=0,1.
    \label{eq:w_alpha_definition}
\end{equation}
The first component is the squared weight on the first site of the associate bond, while the second component is the squared weight on the second site. Hence, for every realization $w_{j,r;1}^{(\alpha)} +w_{j,r;2}^{(\alpha)}=1$.
We now assume that every $\Omega$ is sampled independently from the Haar measure on $O(4)$, for every bond, layer, and time step. The Haar second moment is $\mathbb E_\Omega\left[\Omega_{k\ell}\Omega_{k'\ell'}\right]=\frac14 \delta_{kk'}\delta_{\ell\ell'}$.
It follows that
\begin{equation}
    \mathbb E_\Omega
    \left[
    \bm w_{j,r}^{(\alpha)}
    \right]
    =
    \begin{pmatrix}
        1/2
        \\
        1/2
    \end{pmatrix},
    \qquad
    \alpha=e,o.
    \label{eq:w_alpha_average}
\end{equation}
Equation~\eqref{eq:w_alpha_average} describes the action on a single input
Majorana. For an operator that has already spread over several Majoranas,
one must also verify that the cross terms vanish under the Haar average.

Consider a newly sampled gate acting on the neighbouring sites $(x,x+1)$.
Define $S_x=\{2x,2x+1,2x+2,2x+3\}$ and for every $\ell\in S_x$, the updated coefficient is
\begin{equation}
    O_{a\ell}\left(t+\frac12\right)
    =
    \sum_{k\in S_x}
    O_{ak}(t)\Omega_{k\ell}.
    \label{eq:coefficient_update}
\end{equation}

For one circuit realization, define the squared Majorana weight on site
$x$ by $q_x^{(a)}(t)=\sum_{s=0}^{1} O_{a,\,2x+s}^{\,2}(t)$.
After the new gate,
\begin{equation}
\begin{aligned}
    q_x^{(a)}
    \left(t+\frac12\right)
    &= \sum_{s=0}^{1}
    \left[
    \sum_{k\in S_x}
    O_{ak}(t)
    \Omega_{k,\,2x+s}
    \right]^2
    \\
    &=
    \sum_{k\in S_x}
    O_{ak}^{\,2}(t)
    \sum_{s=0}^{1}
    \Omega_{k,\,2x+s}^{\,2}
    +
    \sum_{\substack{k,k'\in S_x\\k\neq k'}}
    O_{ak}(t)O_{ak'}(t)
    \sum_{s=0}^{1}
    \Omega_{k,\,2x+s}
    \Omega_{k',\,2x+s}.
\end{aligned}
\label{eq:q_half_step_expanded}
\end{equation}
The first term contains the appropriate component of $\bm w_{j,r}^{(\alpha)}$, while the second term contains interference between distinct incoming Majoranas.

Let $\mathcal F_t$ denote the complete random-circuit history before the
new gate is sampled. Conditional on $\mathcal F_t$, the coefficients
$O_{ak}(t)$ are fixed numbers and from the second moment on the Haar ensemble,
\begin{equation}
    \mathbb E_\Omega
    \left[
    \begin{pmatrix}
        q_x^{(a)}\left(t+\frac12\right)
        \\
        q_{x+1}^{(a)}\left(t+\frac12\right)
    \end{pmatrix}
    \,\middle|\,
    \mathcal F_t
    \right]
    =
    \left[
    q_x^{(a)}(t)
    +
    q_{x+1}^{(a)}(t)
    \right]
    \begin{pmatrix}
        1/2
        \\
        1/2
    \end{pmatrix}.
    \label{eq:conditional_vector_rule}
\end{equation}

Define the full ensemble average by $\overline q_x^{(a)}(t)=\mathbb E\left[q_x^{(a)}(t)\right]$, where the expectation includes all gates sampled on all earlier bonds, layers, and time steps. Since the newly sampled matrix $\Omega$ is independent of the history $\mathcal F_t$, we have $\mathbb E[X]=\mathbb E\left[\mathbb E[X\mid\mathcal F_t]\right]$; we obtain
\begin{equation}
    \begin{pmatrix}
        \overline q_x^{(a)}\left(t+\frac12\right)
        \\
        \overline q_{x+1}^{(a)}\left(t+\frac12\right)
    \end{pmatrix}
    =
    \left[
    \overline q_x^{(a)}(t)
    +
    \overline q_{x+1}^{(a)}(t)
    \right]
    \begin{pmatrix}
        1/2
        \\
        1/2
    \end{pmatrix}.
    \label{eq:full_average_vector_rule}
\end{equation}
The conditional average is therefore a recursive evaluation of the full average over the complete spacetime random circuit.

\subsubsection{Odd and even half-step recursions}
For the even layer, the left site of every bond is $x=2m$ for $m=0,...,L/2-1$. Therefore,
\begin{equation}
    \overline q_{2m}^{(a)}
    \left(t+\frac12\right)
    =
    \overline q_{2m+1}^{(a)}
    \left(t+\frac12\right)
    =
    \frac{
    \overline q_{2m}^{(a)}(t)
    +
    \overline q_{2m+1}^{(a)}(t)
    }{2}.
    \label{eq:odd_layer_recursion}
\end{equation}
For the odd layer, the left site of every bond is $x=2m-1$ for $m = 1, ..., L/2-1$. Hence,
\begin{equation}
    \overline q_{2m-1}^{(a)}(t+1)
    =
    \overline q_{2m}^{(a)}(t+1)
    =
    \frac{
    \overline q_{2m-1}^{(a)}
    \left(t+\frac12\right)
    +
    \overline q_{2m}^{(a)}
    \left(t+\frac12\right)
    }{2}.
    \label{eq:even_layer_recursion}
\end{equation}
Using Eq.~\eqref{eq:odd_layer_recursion},
\begin{equation}
    \overline q_{2m-1}^{(a)}
    \left(t+\frac12\right)
    =
    \frac{
    \overline q_{2m-2}^{(a)}(t)
    +
    \overline q_{2m-1}^{(a)}(t)
    }{2}, \quad
    \overline q_{2m}^{(a)}
    \left(t+\frac12\right)
    =
    \frac{
    \overline q_{2m}^{(a)}(t)
    +
    \overline q_{2m+1}^{(a)}(t)
    }{2}.
    \label{eq:odd_input}
\end{equation}
Substitution into Eq.~\eqref{eq:even_layer_recursion} gives
\begin{equation}
\begin{aligned}
    \overline q_{2m-1}^{(a)}(t+1)
    &=
    \overline q_{2m}^{(a)}(t+1)=
    \frac14
    \Big[
    \overline q_{2m-2}^{(a)}(t)
    +
    \overline q_{2m-1}^{(a)}(t)
    +
    \overline q_{2m}^{(a)}(t)
    +
    \overline q_{2m+1}^{(a)}(t)
    \Big].
\end{aligned}
\label{eq:complete_step_q_recursion}
\end{equation}
Defining the total averaged weight on the final even bond by $P_m^{(a)}(t)=\overline q_{2m-1}^{(a)}(t)+\overline q_{2m}^{(a)}(t)$, the odd layer distributes the weight equally between the two sites of each final even bond,
\begin{equation}
    \overline q_{2m-1}^{(a)}(t)
    =
    \overline q_{2m}^{(a)}(t)
    =
    \frac12P_m^{(a)}(t),
    \qquad
    t\geq1.
    \label{eq:q_P_relation}
\end{equation}
Adding the two equal expressions in
Eq.~\eqref{eq:complete_step_q_recursion}, one finds
\begin{equation}
\begin{aligned}
    P_m^{(a)}(t+1)
    =
    \frac12
    \Big[
    &
    \overline q_{2m-2}^{(a)}(t)
    +
    \overline q_{2m-1}^{(a)}(t)
    +
    \overline q_{2m}^{(a)}(t)
    +
    \overline q_{2m+1}^{(a)}(t)
    \Big]=
    \frac14P_{m-1}^{(a)}(t)
    +
    \frac12P_m^{(a)}(t)
    +
    \frac14P_{m+1}^{(a)}(t),
\end{aligned}
\end{equation}
or, equivalently,
\begin{equation}
    P_m^{(a)}(t+1)
    -
    P_m^{(a)}(t)
    =
    \frac14
    \left[
    P_{m-1}^{(a)}(t)
    -
    2P_m^{(a)}(t)
    +
    P_{m+1}^{(a)}(t)
    \right].
    \label{eq:discrete_diffusion_equation}
\end{equation}
This is the discrete diffusion equation for the ensemble-averaged squared Majorana weight on the two-site cells.
Now, we introduce the Fourier transform
\begin{equation}
    P_m^{(a)}(t)
    =
    \int_{-\pi}^{\pi}
    \frac{d\omega}{2\pi}
    e^{i\omega m}
    \widetilde P^{(a)}(\omega,t),
    \label{eq:P_fourier_transform}
\end{equation}
which gives
\begin{equation}
\begin{aligned}
    \widetilde P^{(a)}(\omega,t+1)
    &=
    \left(
    \frac14e^{-i\omega}
    +
    \frac12
    +
    \frac14e^{i\omega}
    \right)
    \widetilde P^{(a)}(\omega,t)=
    \cos^2\left(\frac{\omega}{2}\right)
    \widetilde P^{(a)}(\omega,t),
\end{aligned}
\label{eq:fourier_recursion}
\end{equation}
namely $ \widetilde P^{(a)}(\omega,t)=\cos^{2t}\left(\frac{\omega}{2}\right)\widetilde P^{(a)}(\omega,0)$.
This relation implies that the long-distance, long-time regime is controlled by small momentum $\omega$ and expanding around $\omega=0$ we have
\begin{equation}
\begin{aligned}
    \cos^{2t}\left(\frac{\omega}{2}\right)
    &=
    \left[
    1-\frac{\omega^2}{4}
    +
    O(\omega^4)
    \right]^t\simeq
    e^{-\omega^2t/4},
\end{aligned}
\label{eq:diffusive_fourier_kernel}
\end{equation}
and therefore, in the continuum limit, the cell weight satisfies
\begin{equation}
    \partial_tP^{(a)}(m,t)
    =
    D\,
    \partial_m^2P^{(a)}(m,t),
    \quad
    D
    =
    \frac14 \quad \to \quad P_m^{(a)}(t)
    \simeq
    \frac{1}{\sqrt{\pi t}}
    \exp\left(
    -\frac{m^2}{t}
    \right).
    \label{eq:continuum_diffusion_equation}
\end{equation}
The diffusing object is the ensemble-averaged squared Majorana weight $\overline q_x^{(a)}(t)$, or equivalently the coarse-grained cell weight $P_m^{(a)}(t)$. Hence, the support of the operator in a single circuit realization still expands ballistically inside the circuit light cone.

\subsection{Extension to two and three spatial dimensions}
\label{app:higher_dimensional_diffusion}

The local averaging argument does not depend on the spatial dimension. A
nearest-neighbour Gaussian gate always acts on two complex-fermion sites and
therefore on four Majorana operators. Consequently, the conditional
Haar-averaged rule derived in
Eq.~\eqref{eq:conditional_vector_rule} remains
\begin{equation}
    \begin{pmatrix}
        \overline q_{\mathbf x}^{(a)\prime}
        \\
        \overline q_{\mathbf x+\mathbf e_\nu}^{(a)\prime}
    \end{pmatrix}
    =
    \left[
        \overline q_{\mathbf x}^{(a)}
        +
        \overline q_{\mathbf x+\mathbf e_\nu}^{(a)}
    \right]
    \begin{pmatrix}
        1/2
        \\
        1/2
    \end{pmatrix},
    \label{eq:higher_dimensional_bond_rule}
\end{equation}
where $\mathbf e_\nu$ is a unit vector along one of the lattice spatial directions.
The difference from the one-dimensional case is only in the sequence of
perfect matchings used to cover all nearest-neighbour bonds.

\subsubsection{Two-dimensional circuit}

Consider a periodic square lattice of size $L_x\times L_y$, with $L_x$ and
$L_y$ even. A lattice site is denoted by $\pmb{r}=(x,y)$ with $x = 0,...,L_x-1$ and $y = 0, ..., L_y-1$, and we use the
linear index $\iota_2(x,y)=x+L_x y$, which orders the sites along the $x$ dimension.
The two Majoranas on site $(x,y)$ are therefore
$\gamma_{2\iota_2(x,y)}$ and $\gamma_{2\iota_2(x,y)+1}$.
The evolved Majorana operator is written as
\begin{equation}
    \gamma_a(t)
    =
    \sum_{x=0}^{L_x-1}
    \sum_{y=0}^{L_y-1}
    \sum_{r=0}^{1}
    O_{a,\,2\iota_2(x,y)+r}(t)
    \gamma_{2\iota_2(x,y)+r}.
    \label{eq:2d_majorana_evolution}
\end{equation}
For one circuit realization, the squared Majorana weight per site is
\begin{equation}
    q_{x,y}^{(a)}(t)
    =
    \sum_{r=0}^{1}
    O_{a,\,2\iota_2(x,y)+r}^{\,2}(t),
    \label{eq:2d_q_definition}
\end{equation}
and its full ensemble average is $\overline q_{x,y}^{(a)}(t)=\mathbb E\left[q_{x,y}^{(a)}(t)\right]$.
For the horizontal direction, the circuit geometry is made from the two layers
\begin{equation}
    U_{x,e}
    =
    \bigotimes_{\substack{(x,y)\\x\ {\rm even}}}
    U_{(x,y),(x+1,y)},
    \qquad
    U_{x,o}
    =
    \bigotimes_{\substack{(x,y)\\x\ {\rm odd}}}
    U_{(x,y),(x+1,y)},
    \label{eq:2d_x_layers}
\end{equation}
while for the vertical direction we have
\begin{equation}
    U_{y,e}
    =
    \bigotimes_{\substack{(x,y)\\y\ {\rm even}}}
    U_{(x,y),(x,y+1)},
    \qquad
    U_{y,o}
    =
    \bigotimes_{\substack{(x,y)\\y\ {\rm odd}}}
    U_{(x,y),(x,y+1)}.
    \label{eq:2d_y_layers}
\end{equation}
A complete two-dimensional update is taken to consist of the four
sublayers, applied in the order $ U_{x,e},\qquad U_{y,e}, \qquad U_{x,o},\qquad U_{y,o}$.
All matrices $\Omega$ are sampled independently for every bond, sublayer,
and complete time step.

The averaged action of the horizontal odd and even layers is identical to the one-dimensional calculation at fixed $y$. Likewise, the vertical layers act as the one-dimensional circuit at fixed $x$. Thus, the ordering of the layers does not matter to the averaged weights and we can consider horizontal and vertical sweeps separately. After the complete horizontal application, the averaged weights are equal on the final horizontal bonds, and the vertical application preserves this equality because it acts in the same way on the two members of every horizontal pair. 
It is therefore natural to introduce the four-site cells $C_{m,n}=\left\{\left(
2m-1+\rho_x, 2n-1+\rho_y\right):\rho_x,\rho_y\in\{0,1\}\right\}$ and define the total averaged weight in the cell $C_{m,n}$ by
\begin{equation}
    P_{m,n}^{(a)}(t)
    =
    \sum_{\rho_x=0}^{1}
    \sum_{\rho_y=0}^{1}
    \overline q_{2m-1+\rho_x,\,2n-1+\rho_y}^{(a)}(t).
    \label{eq:2d_P_definition}
\end{equation}
After the first complete two-dimensional update,
\begin{equation}
    \overline q_{2m-1+\rho_x,\,2n-1+\rho_y}^{(a)}(t)
    =
    \frac14
    P_{m,n}^{(a)}(t),
    \qquad
    \rho_x,\rho_y\in\{0,1\}.
    \label{eq:2d_equal_cell_weights}
\end{equation}
The complete horizontal and vertical layers act on the cell weight as a difference
\begin{equation}
    \left(
        \mathcal D_xP^{(a)}
    \right)_{m,n}
    =
    \frac14P_{m-1,n}^{(a)}
    +
    \frac12P_{m,n}^{(a)}
    +
    \frac14P_{m+1,n}^{(a)}, \quad \left(
        \mathcal D_yP^{(a)}
    \right)_{m,n}
    =
    \frac14P_{m,n-1}^{(a)}
    +
    \frac12P_{m,n}^{(a)}
    +
    \frac14P_{m,n+1}^{(a)},
\end{equation}
and the full two-dimensional update is therefore
    $P^{(a)}(t+1)
    =
    \mathcal D_y
    \mathcal D_x
    P^{(a)}(t).$
Explicitly,
\begin{equation}
\begin{aligned}
    P_{m,n}^{(a)}(t+1)
    &=
    \frac14
    P_{m,n}^{(a)}(t)
    +
    \frac18
    \Big[
        P_{m-1,n}^{(a)}(t)
        +
        P_{m+1,n}^{(a)}(t)
        +
        P_{m,n-1}^{(a)}(t)
        +
        P_{m,n+1}^{(a)}(t)
    \Big]
    \\
    &+
    \frac1{16}
    \Big[
        P_{m-1,n-1}^{(a)}(t)
        +
        P_{m-1,n+1}^{(a)}(t)
        +
        P_{m+1,n-1}^{(a)}(t)
        +
        P_{m+1,n+1}^{(a)}(t)
    \Big].
\end{aligned}
\label{eq:2d_exact_update}
\end{equation}
Diagonal terms appear because a complete time step contains both a horizontal and a vertical sweep.
Define the discrete second-difference operators by
\begin{equation}
    \left(
        \Delta_xP
    \right)_{m,n}
    =
    P_{m-1,n}
    -
    2P_{m,n}
    +
    P_{m+1,n},
    \qquad
    \left(
        \Delta_yP
    \right)_{m,n}
    =
    P_{m,n-1}
    -
    2P_{m,n}
    +
    P_{m,n+1}.
    \label{eq:2d_discrete_laplacians}
\end{equation}
Eq.~\eqref{eq:2d_exact_update} can be written as $\mathcal D_x= \mathbbm{1}+\frac14\Delta_x,\,
\mathcal D_y=\mathbbm{1}+\frac14\Delta_y$ and consequently, the exact finite-step equation is
\begin{equation}
\begin{aligned}
    P^{(a)}(t+1)
    -
    P^{(a)}(t)
    &=
    \frac14
    \left(
        \Delta_x
        +
        \Delta_y
    \right)
    P^{(a)}(t)
    +
    \frac1{16}
    \Delta_x\Delta_y
    P^{(a)}(t),
\end{aligned}
\label{eq:2d_exact_difference_equation}
\end{equation}
and mixed term is an exact lattice correction generated by the sequential
composition of the two directional sweeps.
Introduce the Fourier transform
\begin{equation}
    P_{m,n}^{(a)}(t)
    =
    \int_{-\pi}^{\pi}
    \frac{d\omega_x}{2\pi}
    \int_{-\pi}^{\pi}
    \frac{d\omega_y}{2\pi}
    e^{i(\omega_xm+\omega_yn)}
    \widetilde P^{(a)}
    (\omega_x,\omega_y,t),
    \label{eq:2d_fourier_transform}
\end{equation}
then
\begin{equation}
\begin{aligned}
    \widetilde P^{(a)}
    (\omega_x,\omega_y,t+1)
    &=
    \lambda_2(\omega_x,\omega_y)
    \widetilde P^{(a)}
    (\omega_x,\omega_y,t),
    \\
    \lambda_2(\omega_x,\omega_y)
    &=
    \cos^2
    \left(
        \frac{\omega_x}{2}
    \right)
    \cos^2
    \left(
        \frac{\omega_y}{2}
    \right).
\end{aligned}
\label{eq:2d_fourier_eigenvalue}
\end{equation}
At small momenta $\lambda_2(\omega_x,\omega_y)=1-\frac{ \omega_x^2+\omega_y^2 }{4}+ O(\omega^4)$, the mixed lattice term in
Eq.~\eqref{eq:2d_exact_difference_equation} contributes only at fourth
order in momentuma and the long-distance continuum equation is therefore
\begin{equation}
    \partial_tP^{(a)}(m,n,t)
    =
    \frac14
    \left(
        \partial_m^2
        +
        \partial_n^2
    \right)
    P^{(a)}(m,n,t).
    \label{eq:2d_continuum_diffusion}
\end{equation}
For an initially localized cell weight, the asymptotic profile is
\begin{equation}
    P_{m,n}^{(a)}(t)
    \simeq
    \frac{1}{\pi t}
    \exp
    \left[
        -
        \frac{
            (m-m_0)^2+(n-n_0)^2
        }{t}
    \right],
    \label{eq:2d_gaussian_profile}
\end{equation}
with the width along each coordinate that grows proportionally to $\sqrt{t}$.

\subsubsection{Three-dimensional circuit}

The same reasoning follows for the three-dimensional case. The continuum limitn differential equation is therefore
\begin{equation}
    \partial_tP^{(a)}(m,n,p,t)
    =
    \frac14
    \left(
        \partial_m^2
        +
        \partial_n^2
        +
        \partial_p^2
    \right)
    P^{(a)}(m,n,p,t).
    \label{eq:3d_continuum_diffusion}
\end{equation}
For an initially localized cell weight, the asymptotic profile is
\begin{equation}
\begin{aligned}
    P_{m,n,p}^{(a)}(t)
    &\simeq
    \frac{1}{(\pi t)^{3/2}}
    \exp
    \left[
        -
        \frac{
            (m-m_0)^2
            +
            (n-n_0)^2
            +
            (p-p_0)^2
        }{t}
    \right].
\end{aligned}
\label{eq:3d_gaussian_profile}
\end{equation}
In both two and
three dimensions, the diffusion coefficient is $D=1/4$ in coarse-cell
coordinates when one unit of time denotes one complete sequence of all
directional odd and even sublayers.

\section{Entanglement dynamics of Clifford Gaussian brickwork circuits}\label{app:clifford_entanglement}

Consider the same brickwork construction as before, where the two-site gates are now randomly drawn from the Clifford Gaussian set. The linearized action Clifford Gaussian gates over the Majorana operators simplifies to a random permutation: $U^\dagger \gamma_{\pmb{m}} U = \pm \gamma_{\sigma(\pmb{m})}$ with $\sigma \in \mathcal{S}_{2n}$ . This allows to visualize the circuit dynamics in the following way. Starting from the product state $\ket{0}^{\otimes N}$, its fermionic counterpart, i.e. Eq.~\eqref{eq:gamma_zero}, can be viewed  as a series of arcs connecting $\gamma_{(2x-1,y)}$ to $\gamma_{(2x,y)}$. Subsequent applications of random Clifford Gaussian move the arc endpoints to different Majorana sites, widening or twisting the initial arcs. Since the evolution consists only of random permutations, the final state is still described by $N$ arcs connecting the $2N$ Majorana operators. As a consequence, the entanglement at time $t$ can be computed from half the number of arcs crossing the bipartition cut. 

Following Ref.~\cite{paviglianitient2026}, we extend the analysis to two-dimensional circuits. We define $P_t(\pmb{m})=P_t(\pmb{m}|\pmb{m}_0)$, the probability of having an arc endpoint in position $\pmb{m}=(m_x,m_y)$ at time $t$ starting from $\pmb{m}_0=(m_{x_0},m_{y_0})$. As we apply random permutations, the probability of having the arc endpoint at Majorana site $(2x-1,y)$ or $(2x,y)$ is the same at any $t>0$, thus, we redefine $\tilde{P}_t(\pmb{r}) = P(2x-1,y)+P(2x,y)$ for the physical site $\pmb{r}=(x,y)$. 

In the following, we split a single time step of the two-dimensional circuit, which contains four sublayers, into two, such that one contains only the even $x$ and $y$ layers and one contains only the odd ones. Consider being at time $t = t_e$ even. The previous layer is composed of the $U_{x,o}$ and $U_{y,o}$ sublayers; as each of them independently modifies a single coordinate at a time, we can view the probability of the two-dimensional step as the product of the two independent one-dimensional events. If $x$ (or equivalently $y$) is even, then the Majorana in $x$ at time $t$ comes from either position $x$ or $x+1$ at time $t-1$. Instead, the Majorana can reach position $x$ (or $y$) odd,  coming from either $x$ or $x-1$ at time $t-1$. Combining these possibilities, we get
\begin{align}
    \tilde{P}_{t_e}(\pmb{r}) &= \frac{1}{4} [ \tilde{P}_{t_e-1}(2\lfloor x/2\rfloor, 2\lfloor y/2\rfloor)+  \tilde{P}_{t_e-1}(2\lfloor x/2\rfloor+1, 2\lfloor y/2\rfloor) + \nonumber\\
    &+  \tilde{P}_{t_e-1}(2\lfloor x/2\rfloor, 2\lfloor y/2\rfloor+1) +  \tilde{P}_{t_e-1}(2\lfloor x/2\rfloor+1, 2\lfloor y/2\rfloor+1) ] = \tilde{P}_{t_e}(x)\tilde{P}_{t_e}(y).
    \label{eq:odd_prob}
\end{align}
Similarly, if $t = t_o$ is odd, the one-dimensional probability for $s = x,y$ is
\begin{equation}
    \tilde{P}_{t_o}(s) = \frac{1}{2}[ \tilde{P}_{t_o-1}(2\lfloor (s+1)/2\rfloor-1)+  \tilde{P}_{t_o-1}(2\lfloor (s+1)/2\rfloor)]
\end{equation}
leading to $\tilde{P}_{t_o}(\pmb{r}) = \tilde{P}_{t_o}(x)\tilde{P}_{t_o}(y)$.
Each one-dimensional equation gives rise to a binomial probability distribution that spreads over time. Thus, we obtain
\begin{equation}\label{eq:1d_solution}
    \tilde{P}_t(s) =
    \begin{cases}
        \dfrac{1}{2^{t}}\, b\!\left(t-1,\ \dfrac{t-1}{2}
        + \left\lfloor \dfrac{s-1}{2} \right\rfloor
        - \left\lfloor \dfrac{s_0-1}{2} \right\rfloor\right)
        & \text{for odd } t,\\[2ex]
        \dfrac{1}{2^{t}}\, b\!\left(t-1,\ \dfrac{t}{2}-1
        + \left\lfloor \dfrac{s}{2} \right\rfloor
        - \left\lfloor \dfrac{s_0-1}{2} \right\rfloor\right)
        & \text{for even } t,
    \end{cases}
\end{equation}
where $s=x,y$, $s_0 = x_0,y_0$ and $b(n,k)$ is the binomial coefficient $\binom{n}{k}$ for $0\leq k \leq n$ and it is zero otherwise. The product of Eq.~\eqref{eq:1d_solution} for $x$ and $y$ gives the full distribution in $\pmb{r}$, which satisfies the recursion relations with the correct initial condition. At long times, this distribution is well approximated by a Gaussian centered around $\pmb{r}=\pmb{r}_0$ with variance $t$ along each direction, i.e., $P_t(\pmb{r}|\pmb{r}_0) \approx \mathcal{N}(x;x_0,t)\mathcal{N}(y;y_0,t)$.

\begin{figure}
    \centering
    \includegraphics[width=0.8\linewidth]{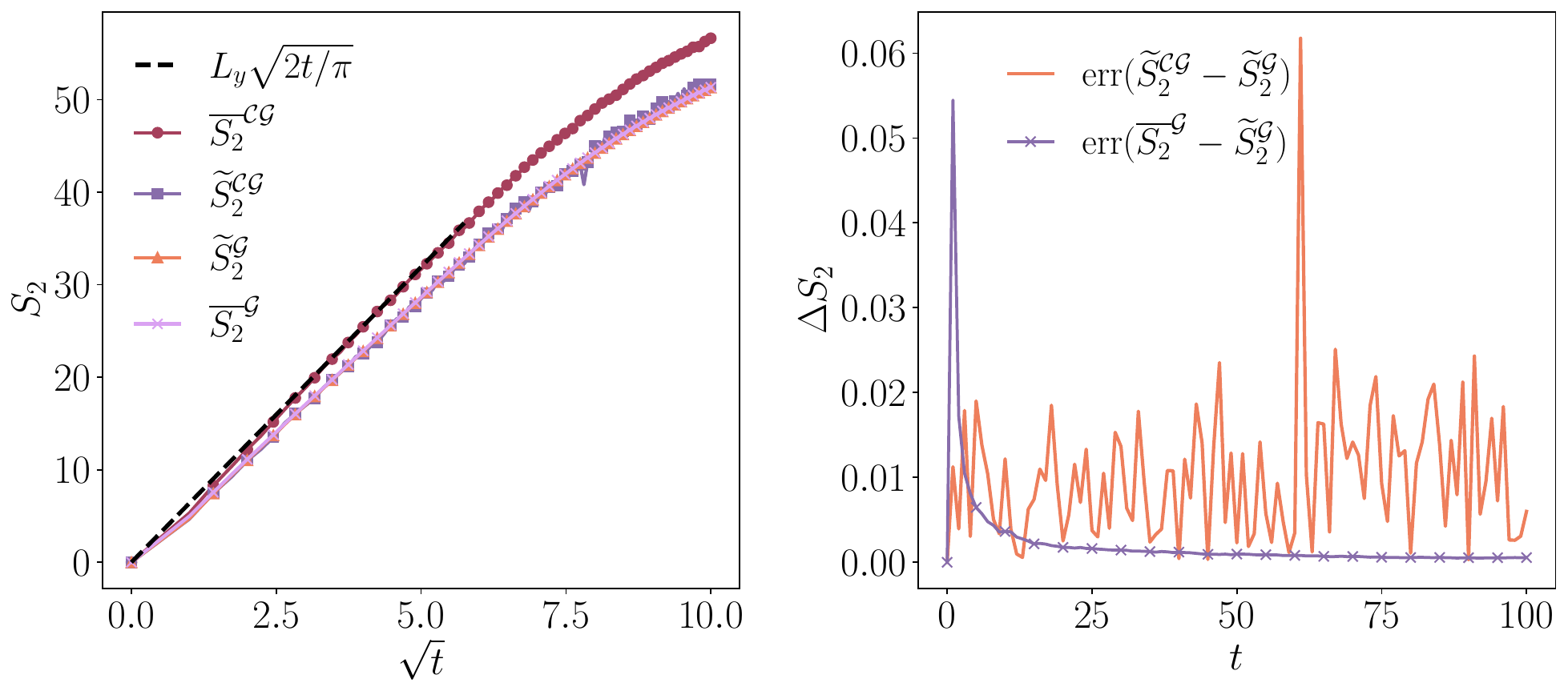}
    \caption{Comparison between different averages of entanglement entropies in a random Clifford Gaussian brickwork circuit (label  $\mathcal{CG}$) and a random Gaussian one (label $\mathcal{G}$). The quenched average $\overline{S}_2^{\mathcal{CG}}$ matches the theoretical prediction of Eq.~\eqref{eq:S2continuum}, while the annealed one $\widetilde{S}_2^{\mathcal{CG}}$ is closely upper-bounded by it. The two Gaussian averages agree with each other and with $\widetilde{S}_2^{\mathcal{CG}}$, following the diffusive growth of entanglement. The right panel represents the relative error of the difference between the matching averages, $\mathrm{err}(x-y) = \frac{x-y}{(x+y)/2}$, where $x,y = \widetilde{S}_2^{\mathcal{CG}}, \widetilde{S}_2^{\mathcal{G}}, \overline{S}_2^{\mathcal{G}}$. Errors remain small except for some points due to numerical precision. Data are taken by sampling $n_{\mathrm{sample}}=250$ times the two random circuits on a grid $32\times 8$ up to depth $t = 100$.}
    \label{fig:clif_comparison}
\end{figure}

Coming to the evaluation of the entanglement entropy, since we consider a vertical cut at physical site $L_x/2$, we need to account for the average number of arcs with an endpoint in $A = \{[1,L_x],[1,L_y]\}$ and the other in $B = \{[L_x+1, 2L_x],[1,L_y]\}$. Moreover, we take the limit of $L_x\gg \sqrt{t}\gg 1$, in which case it is reasonable to assume that the two endpoints of an arc, starting from $(2x_0-1,y_0)$ and $(2x_0,y_0)$ at $t=0$, are independently distributed at $t$. This yields
\begin{equation}
    \overline{S_2}^{\mathcal{CG}}(t) = \frac{1}{2}\sum_{x=1}^{L_x}\sum_{y=1}^{L_y}\left[\sum_{\pmb{m}\in A} P_t(\pmb{m}|(2x-1,y)) \sum_{\pmb{m}'\in B} P_t(\pmb{m}'|(2x,y)) + \sum_{\pmb{m}\in B} P_t(\pmb{m}|(2x-1,y)) \sum_{\pmb{m}'\in A} P_t(\pmb{m}'|(2x,y))\right].
\end{equation}
Decoupling the probabilities and the sums in their $x$ and $y$ components, every $y$ term sums to $\sum_{m_{y}=1}^{L_y}P_t({m_y,y}) = 1 $, and we obtain
\begin{equation}
    \overline{S_2}^{\mathcal{CG}}(t) = \frac{L_y}{2}\sum_{x=1}^{L_x}\left[\sum_{\pmb{m}_x =1}^{L_x} P_t(\pmb{m}_x|2x-1) \sum_{\pmb{m}'_x = L_x+1}^{2L_x} P_t(\pmb{m}'_x|2x) + \sum_{\pmb{m}_x= L_x+1}^{2Lx} P_t(\pmb{m}_x|2x-1) \sum_{\pmb{m}'_x\in 1}^{L_x} P_t(\pmb{m}'_x|2x)\right].
    \label{eq:S2resummed}
\end{equation}
The additional dimension results simply in a prefactor, while the rest of the equation is exactly the one-dimensional one of Ref.~\cite{paviglianitient2026}. Thus, we can take the continuum limit and use the Gaussian approximation of the distribution to simplify Eq.~\eqref{eq:S2resummed} as
\begin{equation}
     \overline{S_2}^{\mathcal{CG}}(t) = L_y\sqrt{\frac{t}{2}}\int_0^u dm_{x_0} [\mathrm{erf}(u+m_{x_0})-\mathrm{erf}(m_{x_0})]
    [\mathrm{erf}(u-m_{x_0})+\mathrm{erf}(m_{x_0})],
    \label{eq:S2continuum}
\end{equation}
where $u = L_x/\sqrt{8t} \gg 1$. The integral can be evaluated explicitly for $u\to \infty$ using that $\mathrm{erf}(u+m_{x_0})\approx 1$ and $\mathrm{erf}(u-m_{x_0})\approx 1$ near $m_{x_0}=0$. Switching back to the time step $t$ used in the Main Text, that includes all four sublayers, the final result at leading order reads $\overline{S_{2}}^{\mathcal{CG}}(t) = L_y \sqrt{2t/\pi}$.
As shown in Fig.~\ref{fig:clif_comparison}, this asymptotic result matches the numerical simulations on Clifford Gaussian circuits for the quenched average $\overline{S_2}^{\mathcal{CG}}$, whereas the annealed average $\widetilde{S}_2^{\mathcal{CG}}$ appears to be smaller, matching the same average, $\widetilde{S}_2^{\mathcal{G}}$, on a generic Gaussian circuit.

\end{document}